\documentclass[%
reprint%
,twocolumn
,amsmath,amssymb
,superscriptaddress
,nofootinbib
,nobibnotes
,a4paper
]{revtex4-1}

\usepackage{graphicx}% Include figure files
\usepackage{xcolor}
\usepackage{bm}% bold math
\usepackage[utf8]{inputenc}
\usepackage[T1]{fontenc}
\usepackage{mathptmx}
\usepackage[cm]{fullpage}

\usepackage[inline]{enumitem}
\usepackage{subcaption}
\newcommand{\noise}{S_{\scriptscriptstyle_\text{Noise}}}

\newcommand{\code}[1]{\texttt{#1}}

\graphicspath{ {./figures/}{./samples/} }

\makeatletter
\renewcommand{\p@subsection}{}
\renewcommand{\p@subsubsection}{}
\makeatother

\begin{document}
\title{Whittle Maximum Likelihood Estimate of spectral properties of
  Rayleigh-Taylor interfacial mixing using hot-wire anemometry experimental data}
\author{David Pfefferlé}
\email{david.pfefferle@uwa.edu.au}
\affiliation{The University of Western Australia, 35 Stirling Highway, Crawley WA 6009, Australia}

\author{Snezhana I. Abarzhi}
\email[corresponding author]{}
\email{snezhana.abarzhi@gmail.com}
\affiliation{The University of Western Australia, 35 Stirling Highway, Crawley WA 6009, Australia}

\begin{abstract}
  Investigating the power density spectrum of fluctuations in Rayleigh-Taylor (RT) interfacial mixing is a means of studying characteristic length- and time-scales, anisotropies and anomalous processes. Guided by group theory, analysing the invariance-based properties of the fluctuations, our work examines raw time series from hot-wire anemometry measurements in the experiment by Akula~\textit{et al.}, JFM~\textbf{816}, 619-660~(2017). The results suggest that the power density spectrum can be modelled as a compound function presented as the product of a power law and an exponential. The data analysis is based on Whittle's approximation of the power density spectrum for independent zero-mean near-Gaussian signals to construct a Maximum likelihood Estimator (MLE) of the parameters. Those that maximise the log-likelihood are computed numerically through Newton-Raphson iteration. The Hessian of the log-likelihood is used to evaluate the Fisher information matrix and provide an estimate of the statistical error on the obtained parameters. The Kolmogorov-Smirnov test is applied to analyse the goodness-of-fit, by verifying the hypothesis that the ratio between the observed periodogram and the estimated power density spectrum follows a chi-squared probability distribution. The dependence of the parameters of the compound function is investigated on the range of mode numbers over which the fit is performed. In the domain where the relative errors of the power law exponent and the exponential decay rate are small and the goodness-of-fit is excellent, the parameters of the compound function are clearly defined, in agreement with the theory. The study of the power-law spectra in RT mixing data suggests that rigorous physics-based statistical methods can help researchers to see beyond visual inspection.
\end{abstract}

 % activate by adding showpacs to documentclass

\maketitle
%

% \begin{quotation}
% The Rayleigh-Taylor instability and associated inter-facial mixing play an important role in a broad range of processes in nature and technology. Examples include supernovae, fusion, fossil fuel extraction and nano-fabrication. Reliable methods of analysis of experimental data are required to better understand Rayleigh-Taylor relevant processes and to achieve a bias-free interpretation of the dynamics. In this paper, we develop the first -- to our knowledge -- formal statistical method of analysis of experimental data from the hot-wire anemometry measurements of Rayleigh-Taylor inter-facial mixing. We apply the method to extract the most likely parameters of a model power density spectrum from the raw time series, and to estimate their statistical errors. The important advantage of our approach is to provide a goodness-of-fit through the systematic testing of residuals. The data analysis results agree with group theory of Rayleigh-Taylor mixing and find that the power density spectrum of experimental quantities is confidently described by the product of a power law and an exponential.
% \end{quotation}

\section{Introduction}
\label{sec:intro}

The Rayleigh-Taylor instability (RTI) develops at the interface between fluids with different densities accelerated against their density gradient~\cite{rayleigh-1883,davies-taylor-1950}. Intense interfacial Rayleigh-Taylor (RT) mixing of the fluids ensues  with time~\cite{rayleigh-1883,davies-taylor-1950,abarzhi-2010a,abarzhi-2010b,meshkov-2006,akula-2017,akula-2016}. Its dynamics is believed to be self-similar~\cite{rayleigh-1883,davies-taylor-1950,abarzhi-2010a,abarzhi-2010b,meshkov-2006,akula-2017,akula-2016}. Particularly in RT mixing induced by constant acceleration, the length scale in the acceleration direction grows quadratically with time~\cite{rayleigh-1883,davies-taylor-1950,abarzhi-2010a,abarzhi-2010b,meshkov-2006,akula-2017,akula-2016}. RTI and RT mixing play important role in a broad range of processes in nature and technology~\cite{abarzhi-2013,arnett-1996,haan-2011}. Examples include supernovae, inertial confinement fusion, material transformation under impact, and fossil fuel extraction~\cite{abarzhi-2013,arnett-1996,haan-2011,peters-2000}. The development of reliable methods of analysis of experimental and numerical data is required to better understand RT-relevant phenomena and to achieve a bias-free interpretation of the results~\cite{abarzhi-2013,anisimov-2013,orlov-2010}.

There are several challenges in studying RTI and RT mixing: the stringent requirements on the flow implementation, diagnostics and control in experiments~\cite{abarzhi-2013,anisimov-2013,orlov-2010,sreenivasan-2018,meshkov-2013,robey-2003,remington-2018}; the necessity to accurately capture interfaces and small-scale dissipation processes in simulations~\cite{ristorcelli-2004,glimm-2013,kadau-2010,youngs-2013}; and the need to account for the non-local, multi-scale, anisotropic, heterogeneous and statistically unsteady character of the dynamics in theory~\cite{abarzhi-2010a,abarzhi-2010b,anisimov-2013,abarzhi-2005}. Furthermore, a systematic interpretation of RT dynamics from data alone is not straightforward and requires a substantial range of highly resolved temporal and spatial scales~\cite{anisimov-2013,sreenivasan-2018}.

Remarkable success was recently achieved in the understanding of the fundamentals of RT mixing~\cite{abarzhi-2010a,abarzhi-2010b,anisimov-2013}.
Particularly, group theory analysis found that symmetries, invariants, scaling and spectral properties of RT mixing may depart from those of isotropic homogeneous turbulence; RT mixing may keep order, due to its strong correlations, weak fluctuations and sensitivity to deterministic conditions~\cite{abarzhi-2010a,abarzhi-2010b,anisimov-2013}. This theory explained experiments, where the order of RT mixing was preserved even at high Reynolds numbers~\cite{akula-2017,meshkov-2013,robey-2003,remington-2018,meshkov-abarzhi-2019}, and simulations, where departures of RT dynamics from canonical turbulent scenario were noted~\cite{ristorcelli-2004,glimm-2013,kadau-2010,youngs-2013}.

An important aspect of RT mixing that requires better understanding is the effect of fluctuations on the overall dynamics~\cite{abarzhi-2010a,abarzhi-2010b,anisimov-2013}.
The appearance of fluctuations in RT flows is usually associated with shear-driven interfacial vortical structures and with broad-band initial perturbations~\cite{abarzhi-2010a,abarzhi-2010b,anisimov-2013,orlov-2010,akula-2017,meshkov-2013,robey-2003,remington-2018}. It is commonly believed that the former may produce small scale irregularities, the latter may enhance the interactions of large scales, and that both may lead RT flow to a self-similar state. Yet, we still need to identify the very nature of fluctuations in RT mixing in order to accurately quantify their properties. By confirming group theory results, experiments~\cite{meshkov-abarzhi-2019} unambiguously found that in a broad range of setups and Reynolds numbers up to $3.2\times 10^6$, the self-similar RT mixing is sensitive to deterministic - the initial and the flow - conditions. We thus need to determine whether in other experiments the fluctuations in RT mixing are chaotic and are set by deterministic conditions, or whether they are stochastic and independent of deterministic conditions.

In this work, the properties of RT mixing are studied through scrupulous analysis of experimental data~\cite{akula-2017}. The data were obtained at the gas tunnel facility~\cite{akula-2017}. The experiments investigated the unstably stratified free shear flows and the coupling of Kelvin-Helmholtz and Rayleigh-Taylor instabilities. One of the 10 setups implemented in these experiments represented 'pure' Rayleigh-Taylor dynamics at Reynolds numbers up to $3.4\times 10^4$~\cite{akula-2017}. Hot-wire anemometry was employed to obtain the fluctuations spectra of the velocity field. The measurements called for a formal analysis of Rayleigh-Taylor experimental data, that would see beyond visual inspection~\cite{akula-2017}.

In this work, our data analysis method is guided by group theory considerations~\cite{abarzhi-2013,anisimov-2013}. Group theory outlines the invariance-based properties of fluctuations, including their spectra and the span of scales~\cite{abarzhi-2013,anisimov-2013,abarzhi-2019}. The resulting empirical model is a combination of power-law and exponential functions which describe the self-similar and scale-dependent parts of the spectrum. We analyse the experimental data represented by raw time series from hot-wire anemometry measurements for the pure Rayleigh-Taylor dynamics in experiments~\cite{akula-2017}. We further choose one of the velocity components which is expected to be the least influenced by the deterministic experimental conditions~\cite{akula-2017}. A formal statistical method is applied to analyse RT mixing data. The method is based on Whittle's approximation of the power density spectrum. It constructs the Maximum Likelihood Estimator (MLE) of the model parameters, numerically solves the optimisation problem through Newton-Raphson iteration algorithm, and estimates statistical errors via the use of the Fisher information matrix obtained from the Hessian of the log-likelihood. The Kolmogorov-Smirnov test~\cite{kolmogorov-1933,smirnov-1948} is further applied to verify the goodness-of-the-fit. We find that, in agreement with the theory, the power density spectrum of experimental quantities can be described by the product of a power law and an exponential. Our work is based on lucid physics background and applies a rigorous statistical technique in order to obtain reliable information from the data describing Rayleigh-Taylor dynamics. Our work is the first (to the authors' knowledge) to provide data analysis at a deeper level than visual inspection, which is traditionally used in experiments and simulations on Rayleigh-Taylor instability and Rayleigh-Taylor interfacial mixing.

\section{Dynamics of self-similar RT mixing}
\subsection{Theory}
\subsubsection{Symmetry and invariance}
\label{sec:theory}
Self-similar RT mixing has a number of symmetries, in a statistical sense, and is invariant with respect to scaling transformations. These symmetries and transformations are distinct from those of canonical Kolmogorov turbulence~\cite{abarzhi-2010a,abarzhi-2010b,anisimov-2013,abarzhi-2019}.
Self-similar canonical turbulence is isotropic and homogeneous; it is inertial and is invariant with respect to Galilean transformations~\cite{kolmogorov-1941a,kolmogorov-1941b,landau-lifshitz-1987}. Self-similar RT mixing is anisotropic and inhomogeneous; it is accelerated and is thus non-inertial and is not Galilean-invariant~\cite{abarzhi-2005,abarzhi-2010a,abarzhi-2010b}.
In canonical turbulence, the invariant quantity of the scaling transformation is the rate of dissipation of specific kinetic energy $\varepsilon \sim v^3/L\sim v_l^3/l$, where $v(v_l)$ is the velocity scale at large (small) length scale $L(l)$~\cite{sreenivasan-2018,kolmogorov-1941a,kolmogorov-1941b,landau-lifshitz-1987}. Its invariance is compatible with the existence of an inertial interval and a normal distribution of velocity fluctuations~\cite{abarzhi-2005,sreenivasan-2018,kolmogorov-1941a,kolmogorov-1941b,landau-lifshitz-1987}. In RT mixing, the invariant quantities of the scaling transformation are the rate of loss of specific momentum $\mu\sim v^2/L \sim v^2_l/l$, along with the rate of gain of specific momentum $\tilde{\mu}\sim g$, in the direction of acceleration with magnitude $g$, with $\tilde{\mu}\sim \mu$, whereas the rate of dissipation (gain) of specific energy is time-dependent $\varepsilon(\tilde{\varepsilon})\sim g^2t$, where $t$ is the time~\cite{abarzhi-2010a,abarzhi-2010b,anisimov-2013,abarzhi-2019}.

\subsubsection{Fluctuations spectra}
In canonical turbulence, the invariance of the energy dissipation rate leads to the spectral density of fluctuations of specific kinetic energy $S(k) \sim \varepsilon^{2/3}k^{-5/3}$ (or $S(\omega)\sim \varepsilon \omega^{-2}$), where $S(k)$ (or $S(\omega)$) is the spectral density and $k(\omega)$ is the wave-vector (frequency). The span of scales is constant $L/l_\nu\sim L(\varepsilon/\nu^3)^{1/4}$ where $l_\nu\sim (\nu^3/\varepsilon)^{1/4}$ is a viscous scale and $\nu$ is a kinematic viscosity~\cite{sreenivasan-2018,kolmogorov-1941a,kolmogorov-1941b,landau-lifshitz-1987}. In RT mixing, the invariance of the rate of momentum loss leads to the spectra for kinetic energy fluctuations, $S(k)\sim \mu k^{-2} (S(\omega)\sim \mu^2\omega^{-3})$ and the span of scales $L/l_\nu\sim t^2(\mu^4/\nu^2)^{1/3}$ growing with time $L\sim \mu t^2$, $l_\nu\sim(\nu^2/\mu)^{1/3}$~\cite{abarzhi-2010a,abarzhi-2010b,anisimov-2013} .

\subsubsection{Sensitivity to deterministic conditions}
In addition to the symmetries, invariances and spectra, an important property of self-similar dynamics is sensitivity of fluctuations to deterministic conditions. In canonical turbulence the invariance of the rate of energy dissipation, $\varepsilon\sim v^3/L\sim v_l^3/l$, leads to diffusion scaling law for velocity fluctuations, with $v\sim T^{1/2}$, $v_l\sim \tau^{1/2}$, where $T$ ($\tau$) is the characteristic time-scale at the large (small) length scale $L$ ($l$). Fluctuations caused by self-similar turbulence are stronger than noise set by deterministic conditions. Canonical turbulence is a stochastic process with no memory of deterministic conditions~\cite{sreenivasan-2018,kolmogorov-1941a,kolmogorov-1941b,landau-lifshitz-1987}.

In RT mixing, the invariance of the rate of momentum loss $\mu\sim v^2/L\sim v_l^2/l$ leads to a ballistic scaling law for velocity fluctuations, with $v\sim T$, $v_l\sim \tau$. Fluctuations caused by self-similar RT mixing are comparable to the noise set by deterministic conditions. RT mixing appears as a chaotic process sensing deterministic conditions~\cite{anisimov-2013,abarzhi-2010a,abarzhi-2010b,abarzhi-2019,meshkov-abarzhi-2019}.

\subsection{Experiment}
\subsubsection{Experiments on the unstably stratified shear flows}
In this work, we consider experimental data from 3-wire anemometry of RT mixing obtained at the multi-layer gas tunnel facility designed to study the unstably stratified shear flows and the coupling between the Kelvin-Helmholtz and Rayleigh-Taylor instabilities~\cite{akula-2017}. The details of the experiments, the diagnostics and the data can be found in~\citet{akula-2017}~(\citeyearpar{akula-2017}). 

In the experiments~\cite{akula-2017}, the fluids with different densities first co-flow in separate channels parallel to one another, in fluid streams with the same or with different speeds, and with the heavy fluid positioned above the light fluid. Next, at the end of the channels, the fluid streams meet, the unstable interface between the fluids forms, the initial perturbation is induced at the fluid interface (by, e.g. a flapping wing), and the fluids enter the tunnel section and start to mix. It is believed that the mixing occurs under the effect of buoyancy, when the speeds of the fluid streams are the same and their densities are distinct, or under the combined effects of the buoyancy and the shear, when the speeds of the streams are distinct.

The experiments~\cite{akula-2017} were conducted over 10 setups, with various values of the fluid density ratio and with various amount of shear of the co-flowing streams. By measuring the mixing width gradient variation along the test section, the experiments suggested that while at early times the flow might be governed by Kelvin-Helmholtz dynamics, at late times the flow might be driven primarily by Rayleigh-Taylor dynamics. The Reynolds number of the fluid mixing was estimated as $\sim 10^3-10^4$ and up to $\sim 3.4\times 10^4$.

\subsubsection{Outline of diagnostics}
For quantifying velocity fluctuations, the experiments~\cite{akula-2017} employed hot-wire anemometry, an experimental technique whereby fine temperature fluctuations are acquired at a fixed (Eulerian) position in a flowing gas stream. The change in resistance of the wire is due to heat exchange with the fluid and is some measure of the flow velocity. For isotropic, homogeneous and statistically steady flows, the measured temperature fluctuations can be viewed as fluctuations of specific kinetic energy of the fluid. This makes hot-wire anemometry a robust and reliable method of diagnostics for canonical turbulence~\cite{orlov-2010,sreenivasan-2018,sreenivasan-1999}. RT mixing is anisotropic, inhomogeneous and statistically unsteady. More caution is required in the interpretation of hot-wire anemometry measurements of RT mixing~\cite{orlov-2010}. To obtain some information on the properties of fluctuations in RT mixing, multiple wires with different orientations can be used to measure temperature (resistance) fluctuations in the direction of acceleration and in the other two transverse directions.

In the experiments~\cite{akula-2017}, for simultaneous measurements of fluctuations of the velocity and the density fields, the hot-wire probe and the cold-wire probe are placed in the flow in close proximity to one another. The hot-three-wire probe with the diameter of $5[\mu m]$ is used to measure the velocity fluctuations, and the cold-wire probe is used to measure temperature. The velocity fluctuations obtained in these measurements depend upon the density fluctuations, with the combined spatial resolution of the measurements evaluated as $\sim 6[mm]$. For the fluctuations of the velocity component, which is normal to
the direction of the acceleration and is also normal to the direction of the co-flowing streams of the fluids, and which is relatively independent of the density fluctuations, especially for fluids with close densities, the spatial resolution is substantially - few fold - higher and is set by a spatial resolution of the probe itself~\cite{akula-2017}. By applying the method of visual inspection, the experiments~\cite{akula-2017} analysed the velocity fluctuations spectra for each of the 10 setups of the unstably stratified shear flows at intermediate and at late times. The spectra were compared with the scaling laws for various buoyant flows and with the $-5/3$ turbulent scaling law. The inertial sub-range was estimated to span one decade or so in the presence of shear.

\subsection{The method and the experimental setup and data}
While the results of the experiments~\cite{akula-2017} are interesting, some fundamental aspects require better understanding. For instance, based on the observations~\cite{akula-2017} at Reynolds numbers up to $\sim 3.4\times 10^4$, one might assume that the late-time dynamics of RT mixing is insensitive to the deterministic conditions. One might further speculate that in the unsably stratified shear flows, the shear serves to transition the flow to a turbulent-like regime and to enlarge the inertial sub-range. One might also try to reconcile these hypotheses with the experimental results~\cite{meshkov-abarzhi-2019} that unambiguously observed the sensitivity of RT mixing to the deterministic conditions at Reynolds numbers up to $\sim 3.2\times 10^6$. To address these issues, a formal physics-based method of analysis of data is required. 

The focus of this paper is on the development of data analysis method. Due to the sensitivity of RT dynamics to the deterministic conditions~\cite{meshkov-abarzhi-2019}, special care is required in choosing the experimental setup and the data set among those of experiments~\cite{akula-2017}. First, to isolate the buoyancy effect from the shear, we need to study the 'pure' Rayleigh-Taylor flow setup. Second, to ensure that the flow is self-similar, we ought to analyse the data taken at the very last times. Third, we have to consider the component of the flow that is expected to be the least affected by the deterministic (the initial and the flow) conditions.

%\subsection{Experimental setup and data}
In the experiments~\cite{akula-2017}, the ‘pure’ Rayleigh-Taylor instability corresponds to the so-called setup A1S0. The data taken at the very late time is presented in~\cite[Figure 22e]{akula-2017}. The velocity component that is the least affected by the flow conditions is the so-called $v$ component. This component of the velocity is chosen as the least-affected, since it is normal to the acceleration and is also normal to the direction of the co-flowing streams of the heavy and the light fluids. The brief outline of the experimental conditions for this particular data set is as follows~\cite{akula-2017}. The density of the heavy (light) fluid is $\rho_{h(l)}=1.18(1.10)[kg/m^3]$ leading to the Atwood number $A=(\rho_h-\rho_l)/(\rho_h+\rho_l)=3.5\times 10^{-2}$ and the effective acceleration $g=Ag_0=3.43\times 10^{-1}[m/s^2]$ where $g_0=9.81[m/s^2]$ is the Earth gravity. The dynamic viscosity of the
heavy (light) fluid is $\mu_{h(l)}=1.83\times 10^{-5}[Pa\cdot s]$ leading to the kinematic viscosity $\nu_{h(l)}=\mu_{h(l)}/\rho_{h(l)} = 1.55(1.66)\times 10^{-5}[m^2/s]$.
This evaluates the wavevector $k_\nu=(g/\nu^2)^{1/3}$ as $k_\nu=(g/\nu^2)^{1/3}=(1.13-1.08)\times 10^3[m^{-1}]$, the viscous length scale $l_\nu=2\pi/k_\nu$ as $l_\nu=2\pi/k_\nu=2\pi(\nu^2/g)^{1/3}=(5.58-5.84)\times 10^{-3}[m]$ and the viscous time scale $\tau_\nu=(g k_\nu)^{-1/2}$ as $\tau_\nu=(g k_\nu)^{-1/2}=(\nu/g^2)^{1/3}\sim(5.09-5.21)\times 10^{-2}[s]$. The scales $k_\nu, l_\nu, \tau_\nu$ are also comparable to those of the mode of fastest growth. The largest horizontal length scale corresponds to $L\sim 31.5 [m]$. The largest vertical length scale is given by the tunnel width $H=1.2[m]$, which is also used to scale the values of lengths. In this data set, the total sampling time is $5\times 10^1[s]$. Time series of the data are acquired at a rate of $10^3[Hz]$ so that the set consists of $5\times 10^4$ data points.

For the 'pure' Rayleigh-Taylor setup A1S0 and for the fluctuations of the $v$ component of the velocity, the spatial resolution is set by the spatial resolution of the probe and is evaluated as $l_{res}=1.26\times 10^{-3}[m]$ with the corresponding dimensionless wavevector $k_{res}=6.00\times 10^3$ \cite{akula-2017}. This indicates that the viscous length scale $l_\nu=(5.58-5.84)\times 10^{-3}[m]$ and the corresponding dimensionless wavevector $k_\nu=(1.36-1.30)\times 10^3$ are well resolved in the experiments \cite{akula-2017}. Note that while the 'pure' RT setup A1S0 and the fluctuations of the $v$ component of the velocity allow one to consider wavevectors up to the resolution limit $k_{res}$, for the experimental data analysed in this paper the signals with dimensionless wavevector values, $k>4.00\times 10^3$ (and with the corresponding length scales $l<1.88\times 10^{-3}[m]$), are interpreted as the instrumental noise. We employ this cautious interpretation for the purposes of consistency, because in other experimental setups in the experiments \cite{akula-2017}, the signals with very high values of the wavevector, $k>4.00\times 10^3$, are interpreted as the instrumental noise.
For further experimental details, the reader is referred to the paper~\cite{akula-2017}.
%Note that for the experimental data analysed in this paper the signals with very high values of the wavevector, $k>4\times 10^3$, are interpreted by the experiment \cite{akula-2017} as the instrumental noise. The wavevector $k\sim 4.00\times 10^{3}$ corresponds to the length-scale $~1.88\times 10^{-3}m$ indicating that the smallest viscous length-scale $l_\nu \sim(2.72-2.85)\times 10^{-3}m$ is resolved in the experiments~\cite{akula-2017}.

\section{Results}
Fitting a theoretical power density spectrum to measurements is usually approached by least-square techniques. The latter may yield some bias when the measurement errors are non-Gaussian. This may happen, for instance, when the data is acquired from complex processes. Maximum-Likelihood Estimators (MLEs) may be used for providing i) an estimate of the model parameters, ii) an estimate of the standard error, iii) a fit rejection criterion to assess the match between observed and theoretical spectra.
Some examples of successful use of the MLEs include: the estimate of the Batchelor cutoff wave-number in temperature gradient spectra of stirred fluid~\cite{ruddick-2000}, peak significance testing in the periodogram of X-ray light curves of active galaxies~\cite{vaughan-2005}, the estimate of dissipation in turbulent kinetic energy in environmental flows~\cite{bluteau-2011} and the spectral power density in other applications~\cite{choudhuri-2004}.

In this work, we develop an MLE-based method to analyse the raw RT data from hot wire anemometry in order to i) estimate the parameters of a theoretical power density spectrum in the form of a power law multiplied by an exponential, ii) estimate the errors on those coefficients and iii) test the statistical relevance of the fitted model against the data.
 
\subsection{Theoretical model of realistic data} From the theoretical point of view, we expect a power-law spectra to be displayed over scales that are far from the largest and smallest scales, and that span a substantial dynamic range~\cite{abarzhi-2010a,abarzhi-2010b,anisimov-2013,sreenivasan-2018,kolmogorov-1941a,kolmogorov-1941b,landau-lifshitz-1987}. For the wave-vectors $k\in(k_{min},k_{max})$, this implies that $\log_{10}(k_{max}/k_{min})\gg 1$ with $k_{min}\gg K$ and $k_{max}\ll k_{\nu}$, where $K\sim L^{-1}, k_\nu\sim l_\nu^{-1}$. Similarly, for the frequency $\omega\in(\omega_{min},\omega_{max})$, this implies that $\log_{10}(\omega_{max}/\omega_{min})\gg 1$ with $\omega_{min}\gg \Omega$ and $\omega_{max}\ll \omega_\nu$, where $\Omega\sim L/v, \omega_\nu\sim l_\nu/v_\nu$. While such conditions are easy to implement in ``mathematical'' fluids, they are challenging to achieve in experiments and simulations, where the values of $K, k_\nu$ are usually finite~\cite{anisimov-2013,akula-2017,sreenivasan-2018,sreenivasan-1999}. Hence, one may expect the spectra to be influenced by processes occurring at scales $\sim K$ and $\sim k_\nu$. The former corresponds to long wavelengths and low frequencies, and is usually associated with the initial conditions and with the effect of slow large-scale processes~\cite{anisimov-2013,sreenivasan-2018,kolmogorov-1941a,kolmogorov-1941b,landau-lifshitz-1987,sreenivasan-1999}. The latter requires more attention, since it is associated with fast processes at small scales. Its influence may lead to substantial departure of realistic spectra from canonical power-laws. In isotropic homogeneous turbulence, these departures are known as anomalous scalings~\cite{sreenivasan-2018,kolmogorov-1941a,kolmogorov-1941b,landau-lifshitz-1987,sreenivasan-1999}.

In experiments, we expect the dynamics to be scale-invariant at scales $k\gg k_\nu$ and be scale-dependent at scales $k \sim k_\nu$~\cite{abarzhi-2010a,abarzhi-2010b,anisimov-2013,sreenivasan-2018,kolmogorov-1941a,kolmogorov-1941b,landau-lifshitz-1987,sreenivasan-1999}. Scale-invariant functions are power-laws and logarithms, and scale-dependent functions are exponentials~\cite{landau-lifshitz-1987}. An empirical function behaving as a power-law $k^\alpha$ for scales $k\gg k_\nu$ and as an exponential $\exp(\beta k)$ for scales $k \sim k_\nu$ is of a compound function $S(k)\sim k^\alpha \exp(\beta k)$ (or $S(\omega)\sim \omega^\zeta \exp(\sigma \omega)$). For turbulent and ballistic dynamics, larger velocities correspond to larger length scales (smaller frequencies). This defines the signs of parameters as $\alpha,\beta < 0$ (or $\zeta,\sigma< 0$).

The compound function $S(k)\sim k^\alpha \exp(\beta k)$ has already been successfully applied in turbulence to describe realistic spectra in experiments and simulations~\cite{kraichnan-1959,sreenivasan-1984,saddoughi-veeravalli-1994,khurshid-2018,sreenivasan-2018}. \citet{kraichnan-1959} derived the compound spectral function for isotropic turbulence at very high Reynolds numbers. \citet{sreenivasan-1984,khurshid-2018} identified experimentally the anomalous behaviour at small scales of the energy dissipation rate in isotropic homogeneous turbulence. The importance of the compound function was recognised for understanding, e.g., passive scalar turbulent mixing \cite{sreenivasan-2018}, turbulent layers~\cite{saddoughi-veeravalli-1994}.

From the physics perspectives, for canonical turbulence, the use of the exponential function in the compound spectrum is justified by the constancy of the scale $k_\nu$, which, in turn, is enabled by the invariance of the energy dissipation rate $\epsilon\sim v^3/L\sim v^3_l/l$ leading to a constant value $k_\nu\sim(\nu^4/\epsilon)^{1/4}$~\cite{kolmogorov-1941a,kolmogorov-1941b,landau-lifshitz-1987,kraichnan-1959}. For Rayleigh-Taylor mixing, upon formal substitution of the time-dependent energy dissipation rate, $\epsilon\sim g^2t$, the value $(\nu^4/\epsilon)^{1/4}\sim (\nu^4/(g^2 t))^{1/4}$ is time-dependent. Remarkably, according to group theory approach~\cite{abarzhi-2010a,abarzhi-2010b,abarzhi-2019,meshkov-abarzhi-2019}, Rayleigh-Taylor mixing is characterised by the invariance of the rate of momentum loss $\mu$, which, in turn, defines the scale $k_\nu$ as $k_\nu \sim (\nu^2/\mu)^{1/3}$. The latter is constant and is set by the acceleration, $\mu\sim \tilde{\mu}\sim g$, and is comparable to the mode of fastest growth $\sim (\nu^2/g)^{1/3}$. Hence, for Rayleigh-Taylor mixing, the use of the scale-dependent exponential function in the compound spectrum is acceptable and physically justified.

Note also that some attempts were recently made to describe turbulent spectra in convective flows, such as Benard-Maragoni and Raleigh-Benard convection, by a stretched exponential function \cite{bershadskii-2019}. Since the (stretched) exponential function is scale-dependent, and since the existence of self-similar dynamics in RT mixing was
unambiguously demonstrated by the experiments at high Reynolds numbers $\sim 3.2\times 10^6$ \cite{meshkov-abarzhi-2019}, we apply here the compound function $S(k)\sim k^\alpha \exp(\beta k)$ to describe the spectral properties of RT mixing.

\subsection{Periodogram smoothing via Whittle MLE (spectrum fitting method)}
\label{sec:data-analysis}
\begin{figure}
  \centering
  \includegraphics[trim={0 0 0 5cm},clip,width=\linewidth]{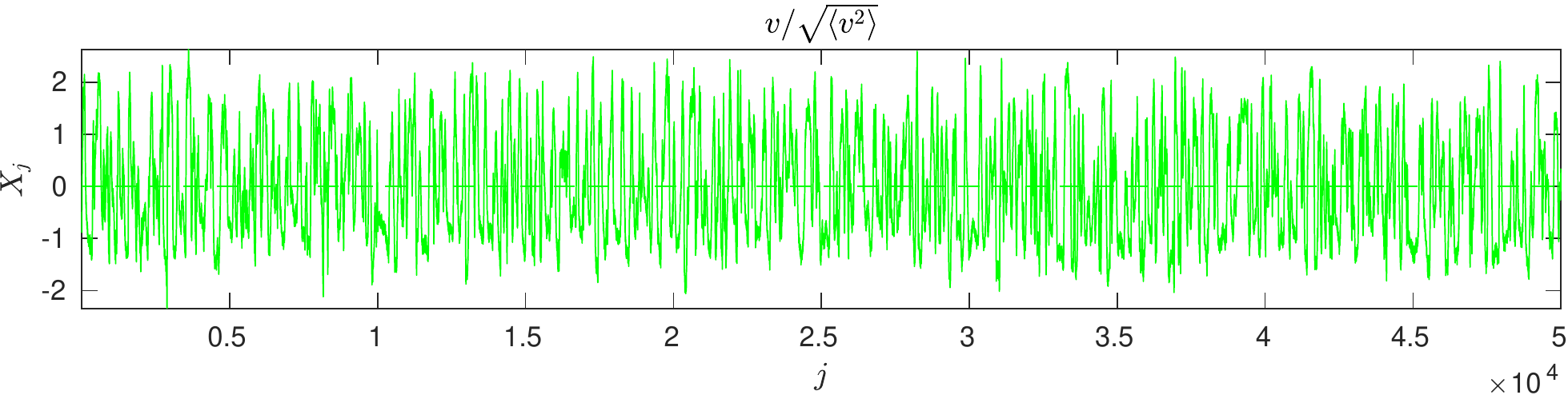}
  \caption{Experimental time series for the normal component $v$ of the flow. The data is normalised by the standard deviation.}
  \label{fig:exp-timeseries}
\end{figure}

The experimental data is given in the form of $N=5\times 10^4$ reals $X_0,\ldots,X_{N-1}$ recorded at constant sampling intervals $\Delta_t=10^{-3}[s]$ in time, see Figure \ref{fig:exp-timeseries}. The signal consists of a zero-mean stationary times series with one-sided power spectral density $S(k)$. In particular, the joint marginal distribution of any part of the series is assumed to be the same as any other part with the same length~\cite{contreras-cristan-2006}.

%\onecolumngrid
% \twocolumngrid

With $N$ even, we compute the discrete Fourier transform (DFT) as the list of complex numbers $\tilde{X}_1,\ldots ,\tilde{X}_{N/2-1}$,
\begin{equation}
  \label{eq:dft}
 \tilde{X}_k = \sum\limits_{j=0}^{N-1} X_j e^{-2\pi i j k/N }.
\end{equation}
The zeroth Fourier coefficient vanishes $\tilde{X}_0=\sum_j^{N-1} X_j=0$ because only the fluctuating part is considered. The periodogram consists of the list of reals $I_k =|\tilde{X}_k|^2/N$, $k=1,\ldots,N/2-1$, see Figure \ref{fig:period_v}.

For an easier graphical comparison between signals with presumably different length-scales, the data is normalised by the standard deviation and the periodogram by the data variance. The latter is motivated by the following property of the DFT,
%\begin{equation}
$ \langle X^2 \rangle \sim \sum_{n=0}^{N-1} \frac{X_n^2}{N} = \sum_{k=0}^{N/2-1}\frac{2I_k}{N} \Rightarrow \bar{I}:=\frac{I}{\frac{N}{2}\langle X^2 \rangle}$.
%\end{equation}
This normalisation is not enforced in the fitting procedure because the true variance is unknown.

We presume that, under suitable conditions~\citep{brillinger-2001}, each Fourier coefficient $\tilde{X}_k$ forms a pair of normally distributed random variables whose variances are approximately equal to the power spectral density, $\text{Var}(\text{Re}(\tilde{X}_k)) = \text{Var}(\text{Im}(\tilde{X}_k)) = S(k)$. The periodogram may thus provide an estimate of the power spectral density such that, for a fixed mode number $k$, the ratio
\begin{align}
 Y_k:=\frac{2 I_k}{S(k)}&\overset{d}{\sim} \chi_2^2,
 & p(I_k) = p_{\chi_2^2}(Y_k) \frac{2}{S(k)} = \frac{e^{-I_k/S(k)}}{S(k)},
\end{align}
is approximately distributed as a chi-square random variable with $2$ degrees of freedom~\cite{brillinger-2001}. Furthermore, the list $\{Y_k\}_{k=1}^{N/2-1}$ forms a collection of (heteroskedastic) random variables that are approximately independent, i.e. $\text{Cov}(Y_k,Y_{k'})\to 0$ as $N\to \infty$ for $k\neq k'$. A Maximum Likelihood Estimator (MLE) is constructed by exploiting the asymptotic behaviour of the periodogram~\cite{whittle-1957} and is based on the following quasi-likelihood function over the range of mode numbers $k=k_l,\ldots,k_r$,
\begin{align*}
% \mathcal{L}(S|X_0,\ldots,X_{N-1})& = \prod_{k=l}^r\frac{e^{-I_k/S(k)}}{S(k)} \\
 \ln\mathcal{L}(S;k_l,k_r|X_0,\ldots,X_{N-1})& = -\sum_{k=k_l}^{k_r}\left[\ln S(k)+\frac{I_k}{S(k)}\right],
\end{align*}
where $k_l$ and $k_r$ are (arbitrary) left and right cutoffs.

As discussed in the foregoing,
% in section \ref{sec:theory}
 we propose to model the RT component of the experimental spectrum in the form of a power law multiplied by an exponential,
\begin{align}
 \label{eq:rtspectrum}
 S_{RT}(k) &= C k^\alpha e^{\beta k} = e^{\alpha \ln k + \beta k + \gamma}. 
\end{align}
We also find useful to account for a low level of instrumental noise~\citep{ruddick-2000}. The power density spectrum is then modelled as
\begin{align}
 \label{eq:modelspectrum}
 S&=S_{RT}+\noise
\end{align}
where the simplest possible noise model is applied. Specifically, a constant white noise of $\noise\sim 10^{-6}-10^{-9}$ mimics the flattening of the periodogram at high mode numbers $k>3000$, see Figure \ref{fig:period_v}.

Our objective is to estimate the three parameters controlling the RT component of the spectrum. Defining the vectors $\vec{\theta}=(\alpha,\beta,\gamma)$ and $\vec{\phi}(k) = (\ln k , k , 1)$, we compute the gradient of the log-likelihood as
\if@twocolumn\begin{multline} \else\begin{equation}\fi
  \partial_{\vec{\theta}} \ln\mathcal{L}(\vec{\theta};k_l,k_r|X_0,\ldots,X_{N-1}) = \if@twocolumn\\\fi
  -\sum_{k=k_l}^{k_r}
  \frac{S_{RT}(k)}{S(k)}\left(1-\frac{I_k}{S(k)}\right)\vec{\phi}(k),
\if@twocolumn\end{multline}\else\end{equation}\fi
as well as the Hessian as
\if@twocolumn\begin{multline} \else\begin{equation}\fi
    \partial^2_{\vec{\theta}\vec{\theta}}\ln\mathcal{L}=:H(\vec{\theta}) = \if@twocolumn\\\fi -\sum_{k=k_l}^{k_r}\frac{S_{RT}(k)^2}{S(k)^2}\left[\frac{I_k}{S(k)}+\left(1-\frac{I_k}{S(k)}\right)\frac{\noise}{S_{RT}(k)} \right]\vec{\phi}(k)\vec{\phi}(k) .
    \if@twocolumn\end{multline}\else\end{equation}\fi
The Maximum likelihood is obtained numerically through a Newton-Raphson method within 6-7 iterations. The scheme and stopping condition are
\begin{equation}
 \hat{\theta}_{i+1} = \hat{\theta}_i - H_i^{-1}\cdot\partial_{\vec{\theta}} \ln\mathcal{L}_i
 \Rightarrow \hat{\theta} \text{ s.t. } \|\partial_{\vec{\theta}}\ln\mathcal{L}(\hat{\theta})\| <\epsilon \sim 10^{-15}
\end{equation}
with the initial condition $\hat{\theta}_0$ obtained via Ordinary-Least-Squares on the log of the periodogram.

\subsection{Error estimation}
The (co)variance on the estimated parameters is bounded from below by the Fisher information matrix of the likelihood function, i.e.
%\begin{equation}
$ \text{Cov}(\hat{\theta},\hat{\theta}) \geq \mathcal{I}^{-1}(\vec{\theta}) \approx -\frac{H^{-1}(\hat{\theta})}{N}$.
%\end{equation}
The error on the model parameters is thus estimated as $\hat{\theta}\pm\sigma$, where $\sigma_i = \sqrt{-H^{-1}_{ii}/N}$. The error is an indicator of the accuracy of the parameter estimation, with more accurate estimations having smaller errors, and with the tolerable accuracy being less that $\sim 10\%$ for relative errors in physics experiments.

\subsection{Goodness-of-fit}
\label{sec:kstest}
We apply the Kolmogorov-Smirnov (KS) test~\citep{kolmogorov-1933,smirnov-1948} to determine whether the alternative hypothesis has statistical significance under the null hypothesis. Our null hypothesis is that the ratio between the observed periodogram and the model power density spectrum is distributed according to a chi-squared distribution with 2 degrees of freedom, $Y_k=\frac{2I_k}{S(k)}\overset{d}{\sim} \chi_2^2$. Departure from the assumed behaviour is detected through the KS test by quantifying the probability, $p_{KS}$, that discrepancies are due only to statistical uncertainty. In the event $p_{KS}$ is too low, the discrepancies cannot be explained by the uncertainty and so the null hypothesis is unlikely (rejected).

In detail, the Empirical Distribution Function (EDF) of the $\eta=k_r-k_l+1$ ordered observations $Y_1<Y_2<\ldots <Y_\eta$,
\begin{equation}
  \label{eq:edf}
 F_\eta(x) = \frac{1}{\eta}\sum_{k=1}^\eta \bm{1}_{(-\infty,x]}(Y_k) \ ,\ \bm{1}_A(z):=\begin{cases}
 1 & z\in A\\
 0 & z \notin A
 \end{cases}
\end{equation}
is compared to the chi-squared Cumulative Distribution Function (CDF) $P_{\chi^2_2}$. The maximum absolute difference between the two distributions,
%\begin{equation}
 $D_\eta := \sup_x|F_\eta(x) - P_{\chi^2_2}(x)|$,
%\end{equation} 
is used as a test statistic. Under the null hypothesis, the value $\sqrt{\eta}D_\eta$ is a random variable distributed asymptotically according to the so-called Kolmogorov distribution~\cite{massey-1951}, i.e.  $\sqrt{\eta} D_\eta \overset{\eta\rightarrow \infty}{\longrightarrow} K$. The null hypothesis is rejected if the distance is larger than the critical value at the significance level $\alpha$. In other words, given a significance level of $\alpha=5\%$ (as per the usual convention), one computes the critical value $K_\alpha$ for which the random variable $\sqrt{\eta} D_\eta$ should remain inferior to in $1-\alpha=95\%$ of the time. If the observed data is such that
\begin{align}
  \label{eq:reject-crit}
 \sqrt{\eta}D_\eta &> K_\alpha, & \Pr(K \leq K_\alpha)=1-\alpha,
\end{align}
then accepting the MLE fit consists of a type II error.

The $p$-value of the test $p_{KS}:=P(K\geq \sqrt{\eta}D_\eta)$ quantifies the probability under null hypothesis of witnessing a discrepancy greater or equal than that observed. A small $p$-value (typically $\leq 5\%$) indicates strong evidence against the null hypothesis, such that the MLE fit must be rejected. A large $p$-value ($>5\%$) indicates weak evidence against the null hypothesis, in which case we fail to reject the MLE fit. We thus interpret a high value of $p_{KS}>5\%$ to indicate a consistent MLE fit. A low value of $p_{KS}$ is interpreted as an inconsistency of the fitting assumptions with the data in regards to
\begin{enumerate*}[label=(\roman*)]
\item the noise model;
\item the left and right cutoffs, $k_l$ and $k_r$;
\item the stationarity of the time series.
\end{enumerate*}
The $p$-value will be quoted in the results as a measure of goodness-of-fit.

The goodness-of-fit (here the Kolmogovor-Smirnov test) is an important part of data analysis in complex systems~\cite{kolmogorov-1933,smirnov-1948}. It ensures that the residuals (i.e. deviations from the fitted spectrum) are distributed in accordance with the assumptions of the fitting technique.
In simple words, when the fit is good the $p_{KS}$ value is high, with a maximum value of $1.00$ ($100\%$), and when the fit is not good, the $p_{KS}$ value is low, with the minimum value of $0.00$ ($0\%$). The threshold $p_{KS}$ value of $0.05$ ($5\%$) is commonly applied in statistics for rejection of a fit.

% In mathematical canonical turbulence the velocity mean is constant, and the velocity fluctuations have Gaussian distribution~\cite{kolmogorov-1941a,kolmogorov-1941b,landau-lifshitz-1987}. The remarkable quality of diagnostics of velocity fluctuations in realistic flows, which was achieved in the past decades, enables the consideration of the velocity fluctuations as yet another 'main signal'. It also requires thorough investigation of departures of data from this signal, that is - the study of fluctuations of the velocity fluctuation in the data set. In this work, we employ the goodness-of-fit and the Kolmogorov-Smirnov test~\cite{kolmogorov-1933,smirnov-1948} in Rayleigh-Taylor interfacial mixing in order to analyse the departures of realistic experimental data~\cite{akula-2017} from the compound function fit of the power-density spectrum of the velocity flucutations and to ensure that the distribution function of these departures is statistically confident and well-defined.

\subsection{Effect of range of wavevector values}
The compound function $S(k)\sim k^\alpha\exp(\beta k)$ is a product of a power-law and an exponential. The exponential $\exp(\beta k)$ is scale-dependent, and its scale $\sim|\beta|^{-1}$ establishes the natural range of $k$ values for the function evaluation. The power-law $k^\alpha$ is scale-invariant, and a substantial span of $k$ values is required for the function evaluation. In the data set considered in our paper, the range of $k$ values is relatively short. This can influcence the fitting of the parameters of the compound function $S(k)\sim k^\alpha\exp(\beta k)$. Hence there is a need to quantify the effect of the left and right cut-offs $k_l$ and $k_r$ delimiting the range of the values $[k_l,k_r]$ and the fitting interval, which are included to determine the parameters of the compound function. Note that this effect commonly exists in turbulent flows. In our work, we study the dependence of the compound function parameters, including the exponent of the power law $\alpha$ and the length-scale of the exponential $|\beta|^{-1}$, on the left and right cut-offs, $k_l$ and $k_r$ and the range of values $[k_l,k_r]$ over which the fit is performed.

\section{Properties of RT data}
\subsection{Spectral properties of experimental data}
\begin{figure}
 \centering
 \includegraphics[width=\columnwidth]{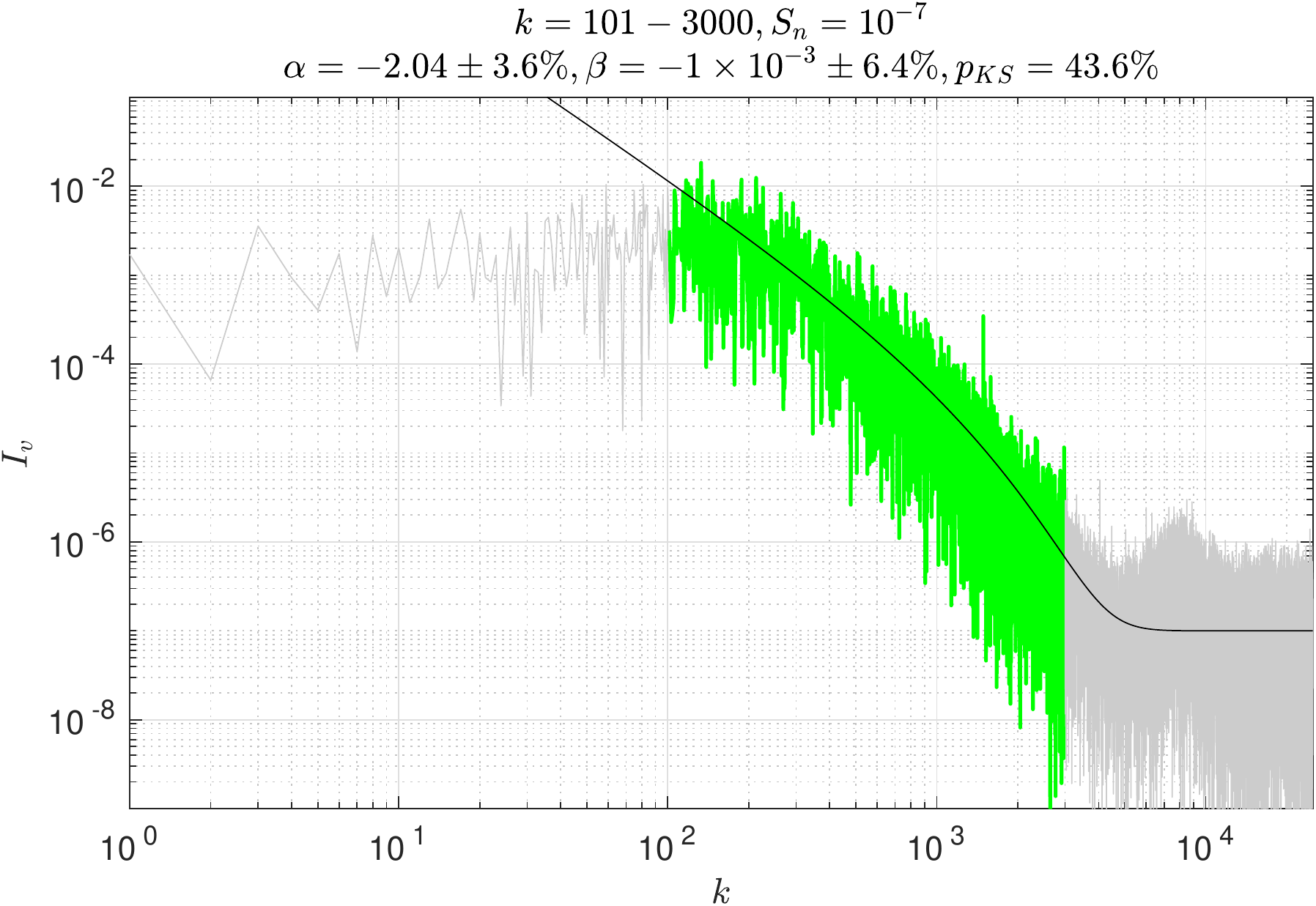}
 \caption{Periodogram of the $v$ components of the RT flow rate on a
   log-log scale (normalised by the data variance). The black line is
   the resulting fit from the MLE of the power law and exponential
   decay rate coefficients $\alpha$ and $\beta$ over a broad range of
   mode numbers $k=101-3000$, assuming a noise level of
   $\noise=10^{-7}$. Grey-coloured data points were excluded from the
   fit.}
 \label{fig:period_v} 
\end{figure}
Figure \ref{fig:period_v} shows the periodogram computed from the times series in Figure \ref{fig:exp-timeseries} characterising fluctuations in the normal component $v$ of the flow. The black line in this Figure represents the model power density spectrum, whose parameters are adjusted through the fitting procedure described in previous sections. % \ref{sec:data-analysis}.
For this particular MLE fit, a broad range of mode numbers $k=101-3000$ is selected. The ``active'' modes are highlighted by colouring the data in green. The grey data depicts the amplitudes of the periodogram that have been excluded from the fit. The end ``tail'' of the periodogram beyond $k>4000$ predominantly reflects instrumental noise. The fitted power density spectrum becomes flat above $k>4000$ due to the choice of noise level at $\noise=10^{-7}$. This adjustable parameter acts effectively as a cutoff by reducing the importance of amplitudes below $\noise$ in the MLE. $\noise$ is selected to be visually consistent with the data.

At high mode numbers, an exponential behaviour is revealed by the non-vanishing $\beta$ coefficients. This exponential behaviour is even more prominent from the strong linear correlation between $k$ and $\ln S_{RT}(k)\sim \beta k $ on a lin-log scale (not shown here). The associated characteristic length-scale corresponds to $k\sim 1/\beta \sim 1000$, which is comparable to $k_\nu$, and which is much lower than both the right fitting limit $\sim 3000$ and the instrumental noise beyond $k>4000$. This indicates that the corresponding length-scale is a physical feature of the flow.

At low mode numbers, the power-law dominates over the exponential term. The fitted power density spectrum thus approaches a line with slope $\alpha$ in log-log scale. Below $k<70$ however, the fit  overestimates the periodogram, as seen through the rise of the black line well above the data points. This departure can be interpreted in several ways. Possible interpretations of this departure may include the statistical unsteadiness of the flow, the effects of the largest vertical and the largest horizontal scales, the scale-dependent dynamics at very large length scales, etc. This behaviour is consistent with sensitivity of the dynamics to the initial conditions at very large length scales (small mode numbers) and with the exponential character of spectra in deterministic chaos.

The error estimation of the parameters $\alpha$ and $\beta$ of the compound function is discussed in more detail below. Briefly, the parameters of the compound function can be accurately identified over a broad range of intervals $[k_l,k_r]$. In order to accurately estimate the power-law exponent $\alpha$ describing the left part of the spectrum, $k\ll k_\nu$, one should account for the significant number of modes on the right-end ($k_r$) of the periodogram. In order to accurately estimate the exponential decay rate $\beta$ desribing the right part of the spectrum, $k\sim k_\nu$, one should account for the significant number of modes on the left-end ($k_l$) of the periodogram.

The effect of the range of values $[k_l,k_r]$ and the left and right cut-offs $k_l$ and $k_r$ on the parameter estimation is discussed in more detail below.
% in section \ref{sec:scans}
Briefly, the $\alpha$ coefficient from the MLE fit becomes smaller in absolute value when the left limit is lowered. This may mean that the power-law loses relevance when the mode number range is extended very far to the left, since the flattening at low mode numbers can be achieved by the exponential term alone. The goodness-of-fit however worsens as we lower the left limit $k_l$. The fit must actually be rejected below $k_l<30$ for $v$ with fixed $k_r=3000$.

An equivalent MLE fitting procedure can be applied to verify whether the periodogram can be described only by an exponential term or only by a power-law. The exponential fit presumes a scale-dependent dynamics, and RT mixing is self-similar. The power-law fit presumes a self-similar dynamics displayed over scales spanning a substantial dynamic range. The power-law fit is discussed in detail below. Briefly, while some reasonable parameter estimations might result from this procedure, the goodness-of-fit suggests that the dynamics is characterised by a power density spectrum that is at least as complicated as the compound function represented by the product of a power-law and an exponential decay over a broad range of scales.

\subsection{Analysis of residuals and goodness-of-fit}
The procedure described above
% in section \ref{sec:kstest}
is applied to the MLE fit reported in Figure \ref{fig:period_v} to assess the goodness-of-fit. The KS test returns a $p$-value of $p_{KS}=43.6\%>5\%$. The probability of witnessing a greater discrepancy between the fit and the data through statistical uncertainty is larger than the adopted rejection level of $5\%$; we interpret the MLE fit as being consistent/valid.

\begin{figure}
  \centering
  \includegraphics[width=\columnwidth]{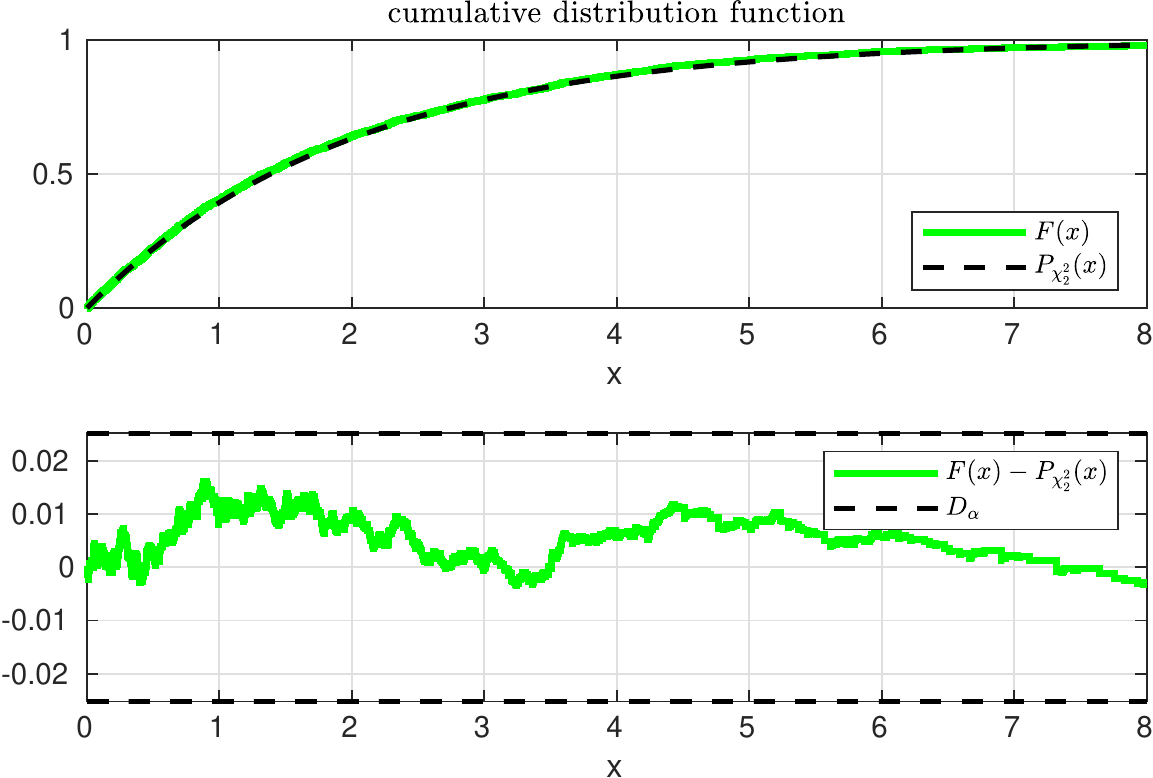}
  \caption{KS test comparing empirical cumulative distribution function of $Y_k=2I_k/S(k)$ with chi-square CDF for MLE fit of Figure \ref{fig:period_v}. The dashed black lines in the second row highlight the critical value of the absolute maximum difference at $5\%$ significance level. The $x$-axis is the upper bound of the interval $(-\infty,x]$ such that $F_\eta(x)$ represents the number of elements in the sample $\{Y_k\}_{k=k_l}^{k_r}$ whose value is smaller or equal to $x$, as per equation (\ref{eq:edf}).}
 \label{fig:ksv}
\end{figure}
The top plot of Figure \ref{fig:ksv} shows the details of the goodness-of-fit procedure by comparing the empirical cumulative distribution function of the collections of ratios $Y_k=2I_k/S(k)$ (solid coloured curve) and the chi-squared CDF (dashed black curve). The difference between the graphs is almost imperceptible. The coloured curve on the bottom plot of Figure \ref{fig:ksv} is the absolute maximum difference between the empirical and chi-squared CDF and the dashed line represents the critical value from the KS statistics $D_{5\%}\sim 2.5\times 10^{-2}$ beyond which the MLE fit must be rejected. As seen on Figure \ref{fig:ksv}, the dashed line is not exceeded.

\begin{figure}
  \includegraphics[width=\columnwidth]{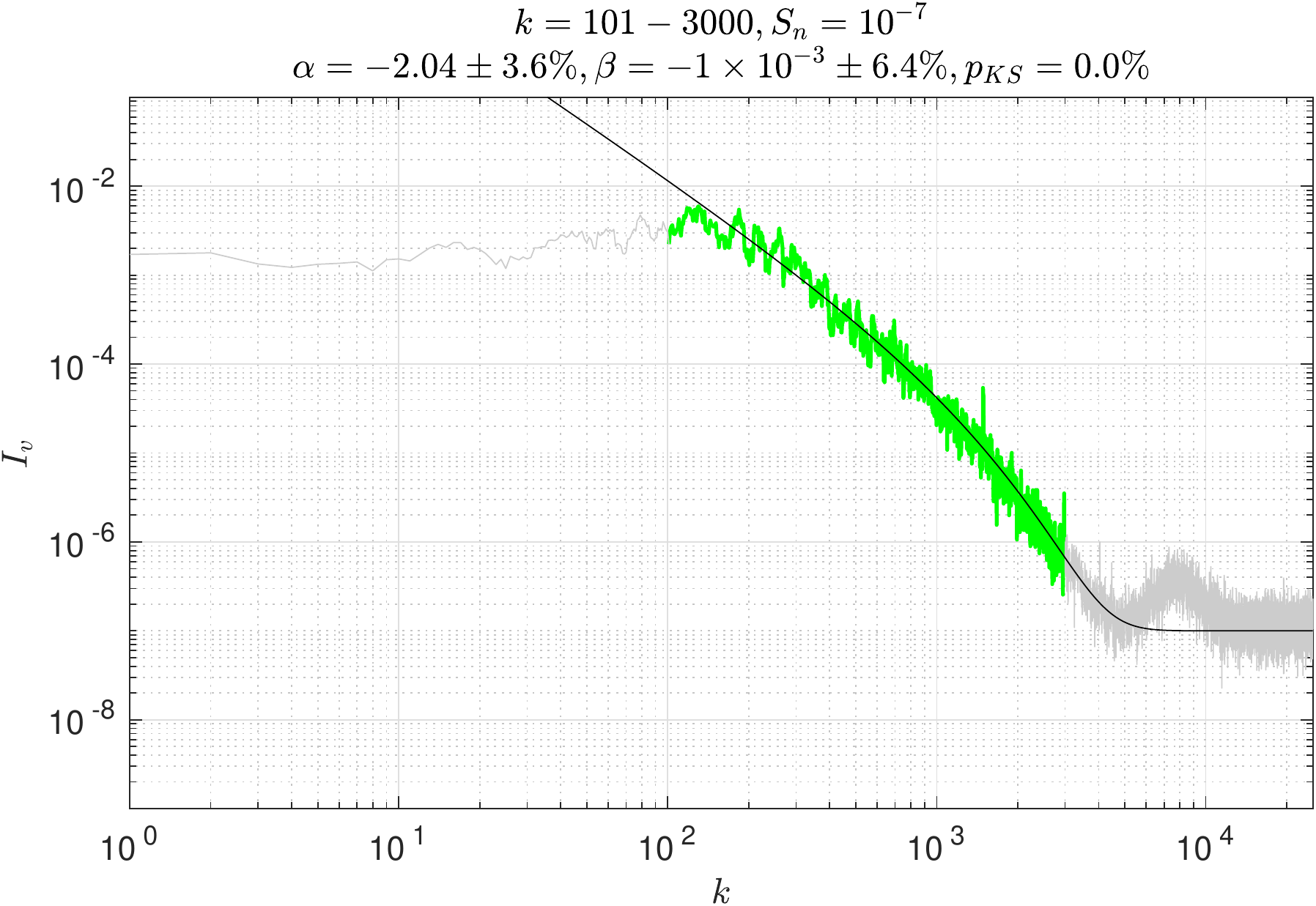}
  \caption{Identically to Figure \ref{fig:period_v}, the MLE is
    applied to MATLAB ``smoothed'' periodogram with little
    difference in the resulting parameter estimation. The KS test
    however severely invalidates the fit, as demonstrated in Figure
    \ref{fig:kssmo_v}.}
  \label{fig:smo_v}
\end{figure}
To highlight the critical importance and sensitivity of the KS test as a rejection method, an identical MLE fit, based on the mode number range $k=101-3000$ and with noise level $\noise=10^{-7}$, is repeated after applying MATLAB's \code{smooth} function to the periodogram, which uses a moving average method with a span of 9 points. Figure \ref{fig:smo_v} shows the processed periodogram as in~\cite{akula-2017} as well as the resulting MLE fit (black curve). As a result of the smoothing, the contours of the periodogram are much cleaner and more precise. The features already present in Figure \ref{fig:period_v} become more prominent; the signals exhibit exponential decay at large mode number before being dominated by a flat spectrum of instrumental noise. An additional interesting feature appears in the very high-end of the spectrum around $k\sim 5000$, namely the presence of a local maximum. A more complicated noise model and/or recording the instrumental noise without flows would be required to model this bump. The detailed analysis of $\noise$ is of minor interest and has little impact on the parameter estimation, knowing that the right mode number limit never exceeds $r<3000$ in our scans.

The parameter estimation resulting from the MLE fit of the periodograms in Figure \ref{fig:smo_v} is exactly the same as for Figure \ref{fig:period_v}, namely $\alpha=-2.04 \pm 3.6\%$ and $\beta=-1\times 10^{-3}\pm 6.4\%$ for the $v$-component of the velocity. Nevertheless, the smoothing adds no benefit to the fitting. To the contrary, it completely ruins the goodness-of-fit.
\begin{figure}
  \includegraphics[width=\columnwidth]{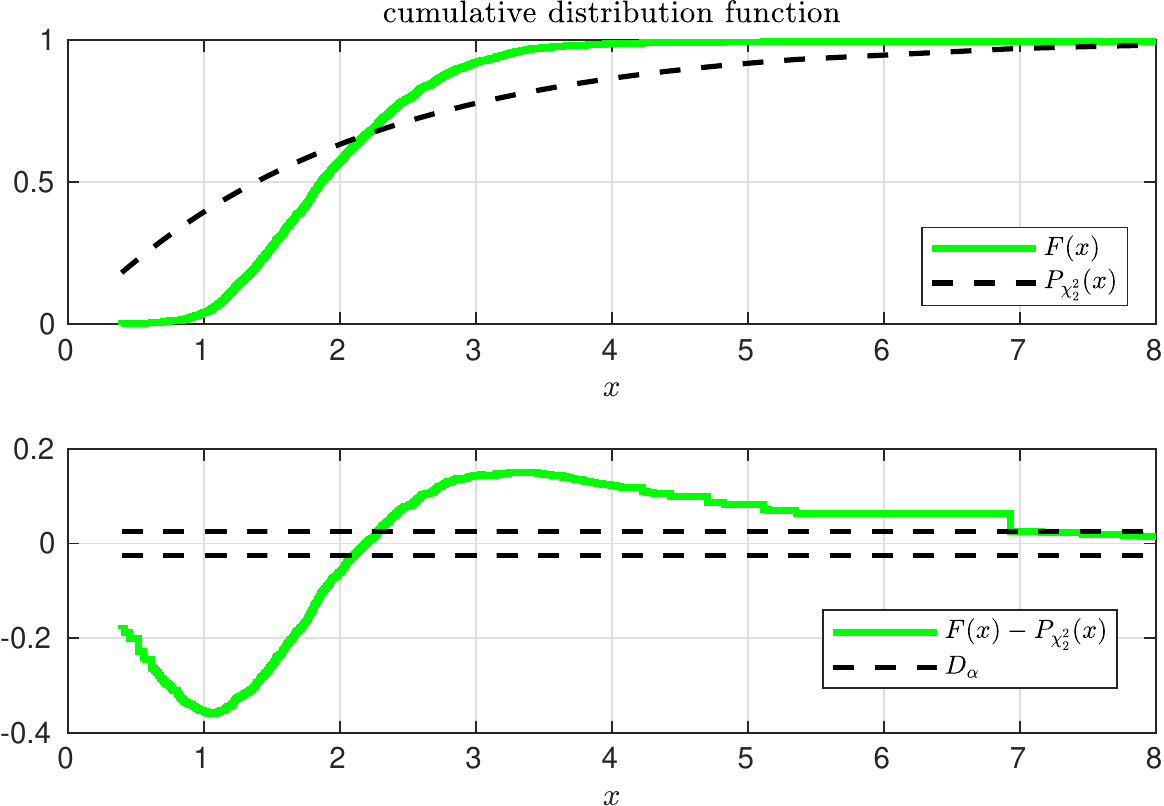}
    \caption{Same caption as Figure \ref{fig:ksv}. KS test rejects the MLE fit of the ``smoothed'' periodogram.}
    \label{fig:kssmo_v}
\end{figure}
The KS test reveals that the residuals have lost consistency with the fitting assumptions. The top plot of Figure \ref{fig:kssmo_v} compares the empirical CDF of the ratios $Y_k=2I_k/S(k)$ (solid coloured curve) with the chi-squared CDF (dashed black curve). In this case, the deviation between the CDFs is clear. The bottom plot of Figure \ref{fig:kssmo_v} show that the critical rejection distance $D_{5\%}$ is exceeded for almost all data points. The important conclusion from this consideration is that data processing is unnecessary prior to applying the MLE fitting procedure and working with raw data is preferable even though noisier.

\subsection{The effect of the range of wavevector values and the left and right cut-offs}
\label{sec:scans}
In this section, we thoroughly investigate the dependence of the MLE fit on the range of values $[k_l,k_r]$, particularly on the left and right cutoffs, $k_l$ and $k_r$ respectively. This study is necessary since the compound function $S(k)\sim k^\alpha\exp(\beta k)$ has the scale-invariant power-law component, and since the range of values of $k$ of the data set is relatively short. To our knowledge, such analysis has never been performed before.

\begin{figure}
  \begin{subfigure}[t]{0.48\textwidth}
    \includegraphics[width=\columnwidth]{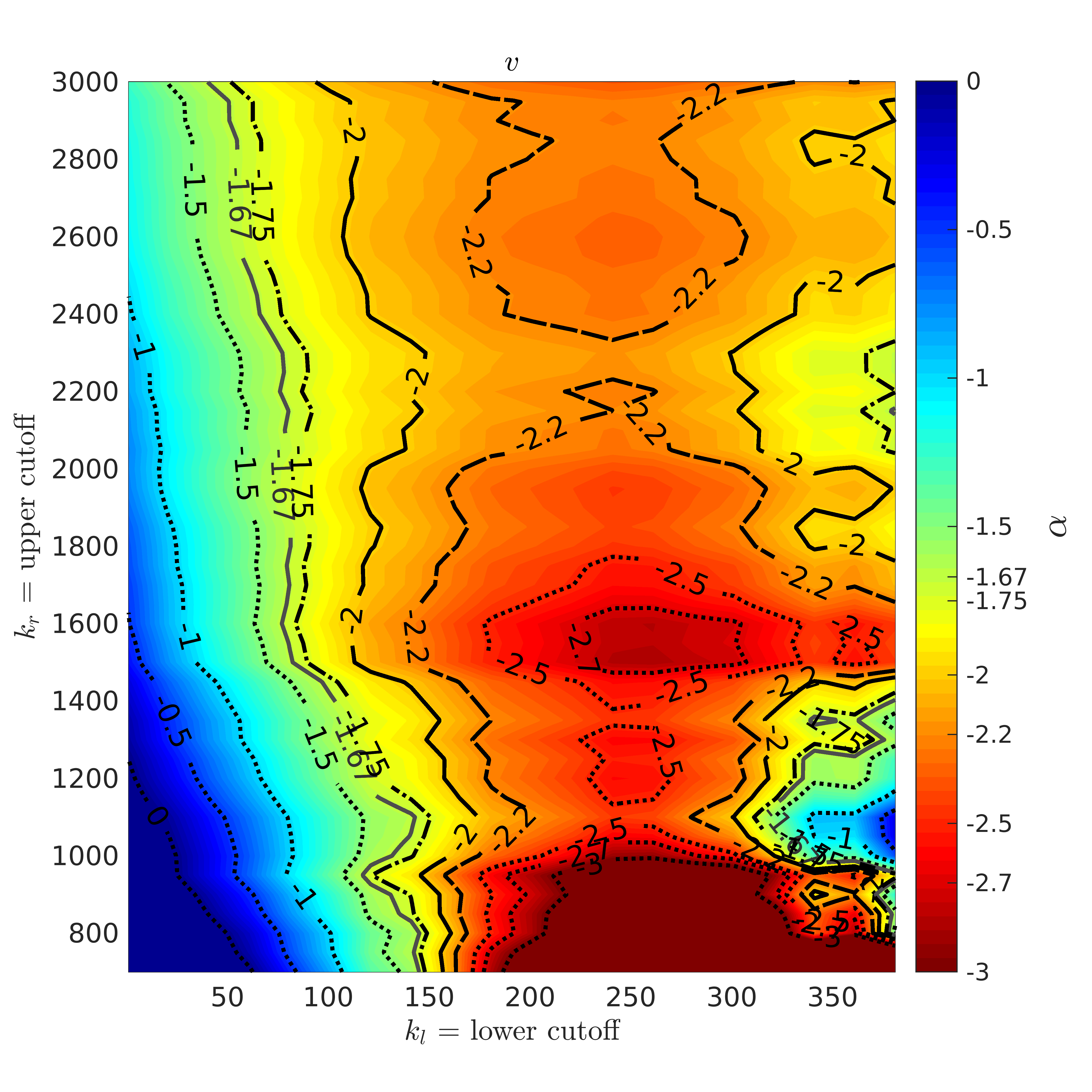}
    \caption{Power law exponent $\alpha$.}
    \label{fig:scan_v_alpha}
  \end{subfigure}
  \begin{subfigure}[t]{0.48\textwidth}
    \includegraphics[width=\columnwidth]{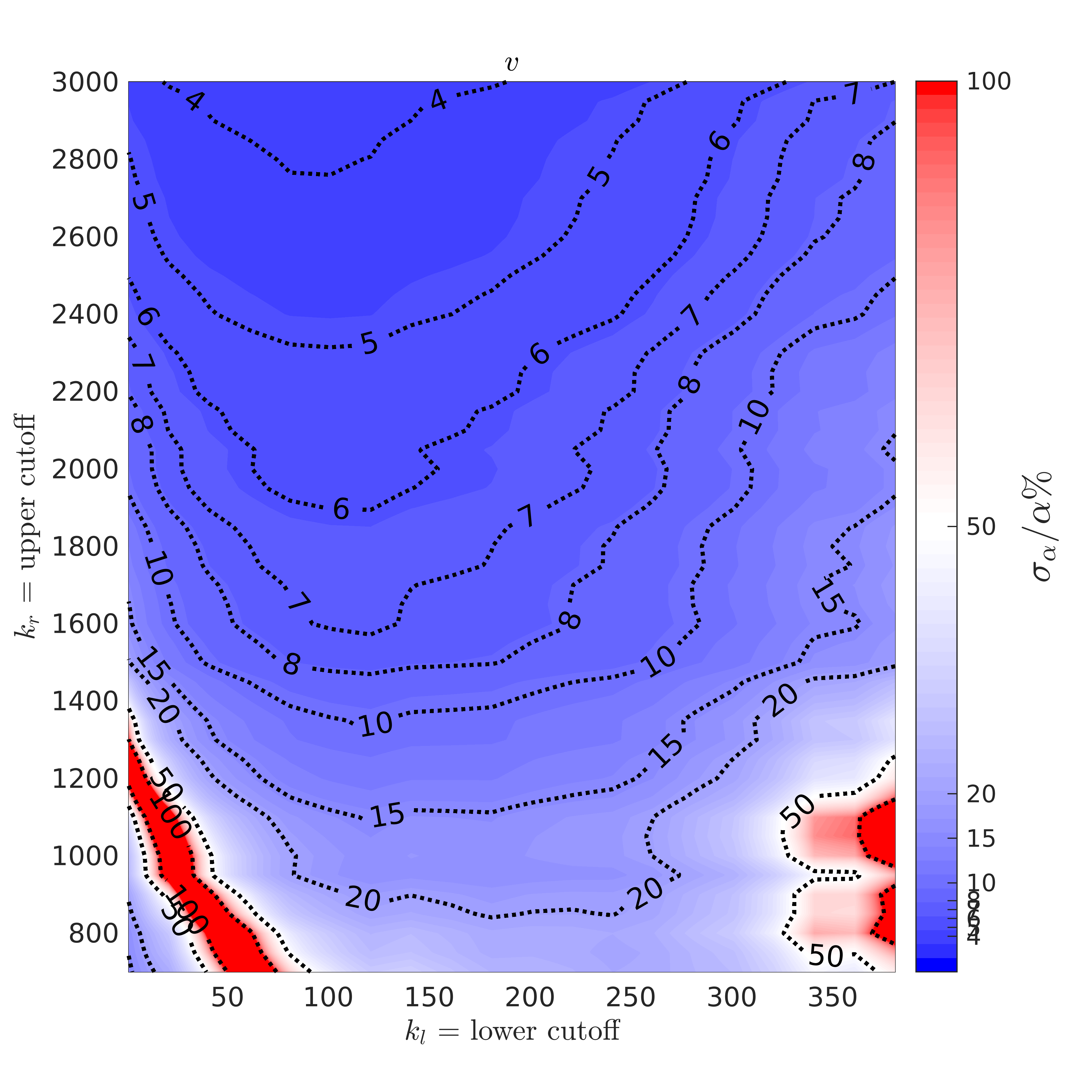}
    \caption{Relative error $\sigma_\alpha/\alpha$ in percent.}
    \label{fig:scan_v_erra}
  \end{subfigure}
  \caption{Dependence on the left and right window limits $k_l$ and $k_r$ of the estimated fiting parameter $\alpha$ and its relative error $\sigma_\alpha/\alpha$.}
\end{figure}

\begin{figure}
  \begin{subfigure}[t]{0.48\textwidth}
    \includegraphics[width=\columnwidth]{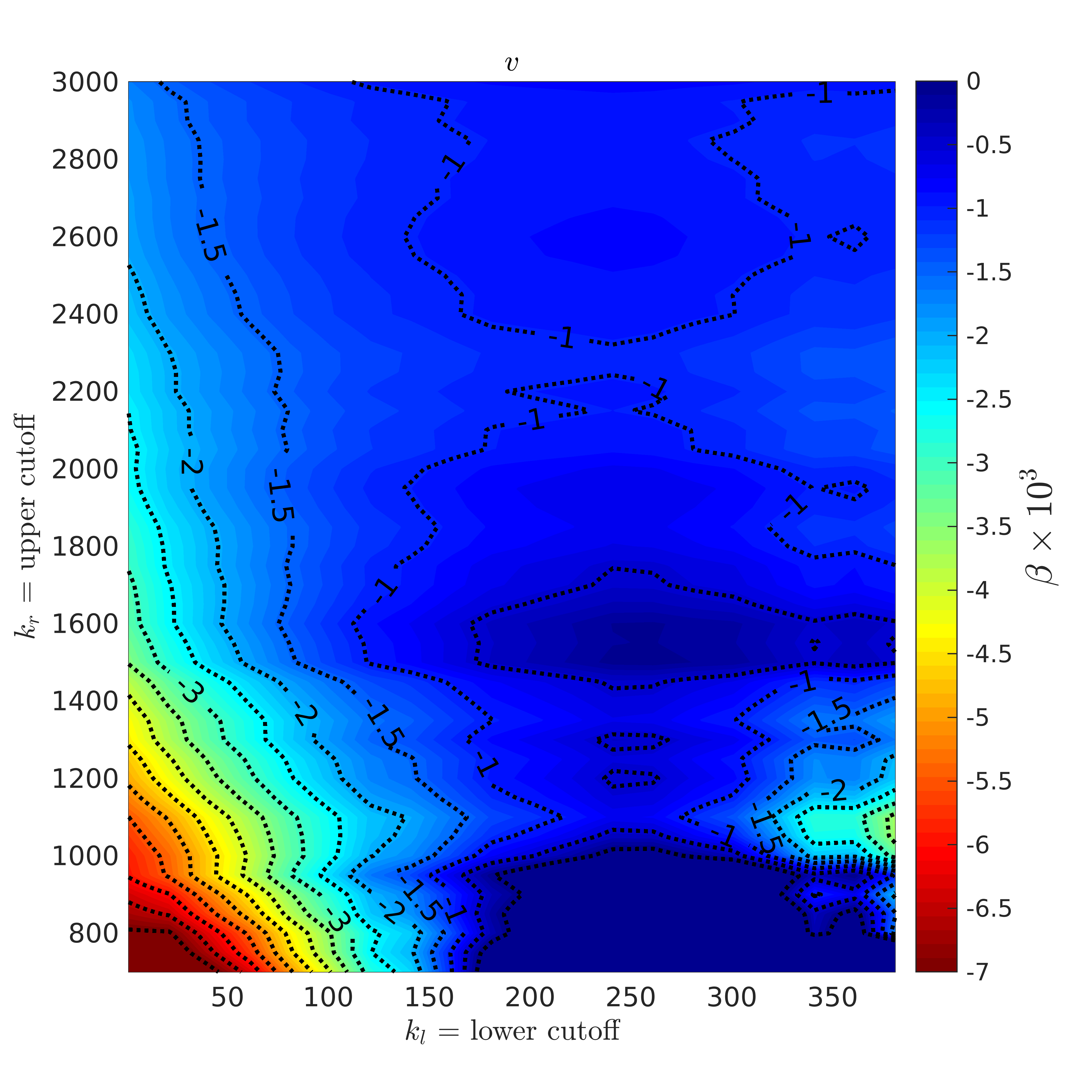}
    \caption{Exponential decay coefficient $\beta$.}
    \label{fig:scan_v_beta}
  \end{subfigure}
  \begin{subfigure}[t]{0.48\textwidth}
    \includegraphics[width=\columnwidth]{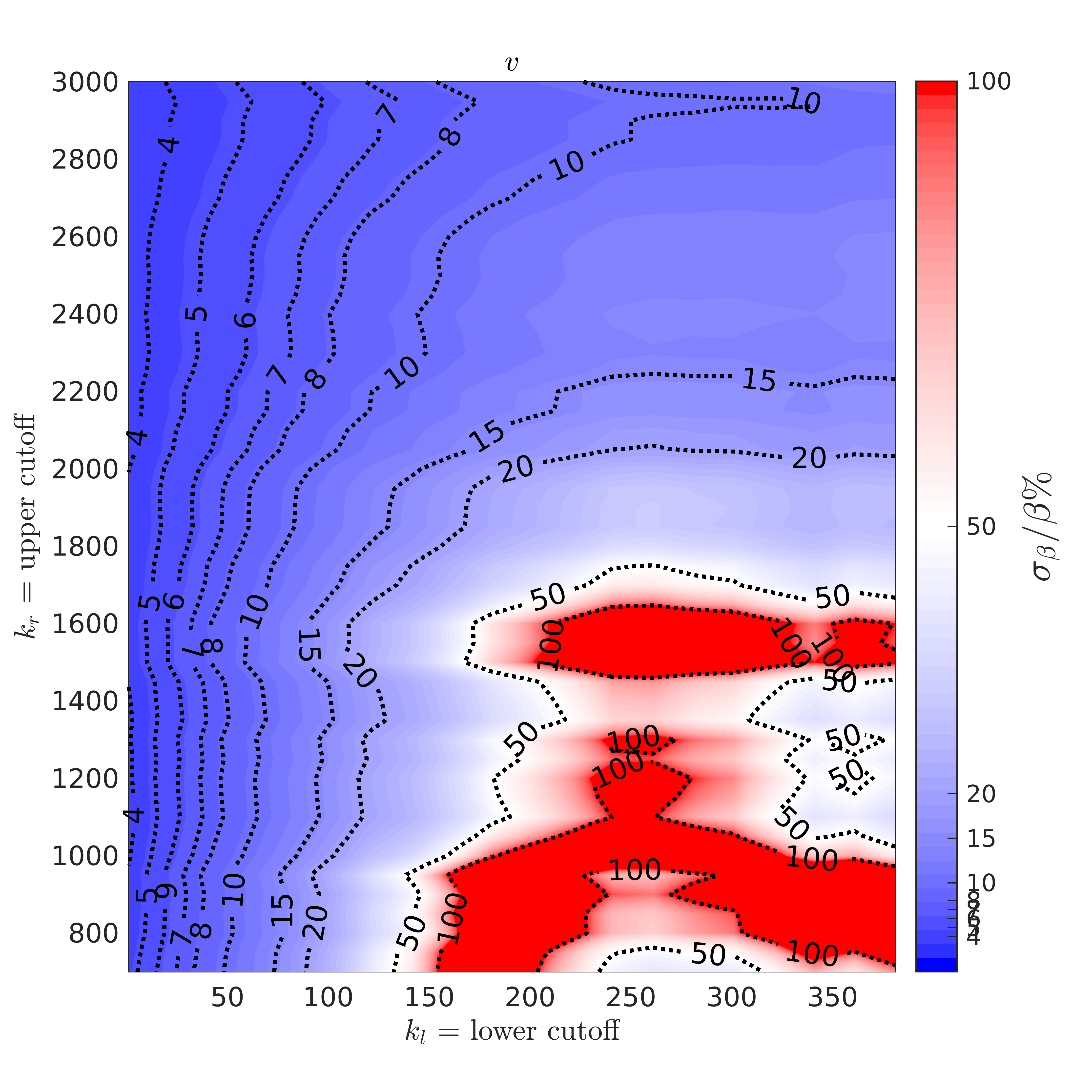}
    \caption{Relative error $\sigma_\beta/\beta$ in percent.}
    \label{fig:scan_v_errb}
  \end{subfigure}
    \caption{Dependence on the left and right window limits $k_l$ and $k_r$ of the estimated fiting parameter $\beta$ and its relative error $\sigma_\beta/\beta$.}
  \end{figure}

  \begin{figure}
    \includegraphics[width=0.48\textwidth]{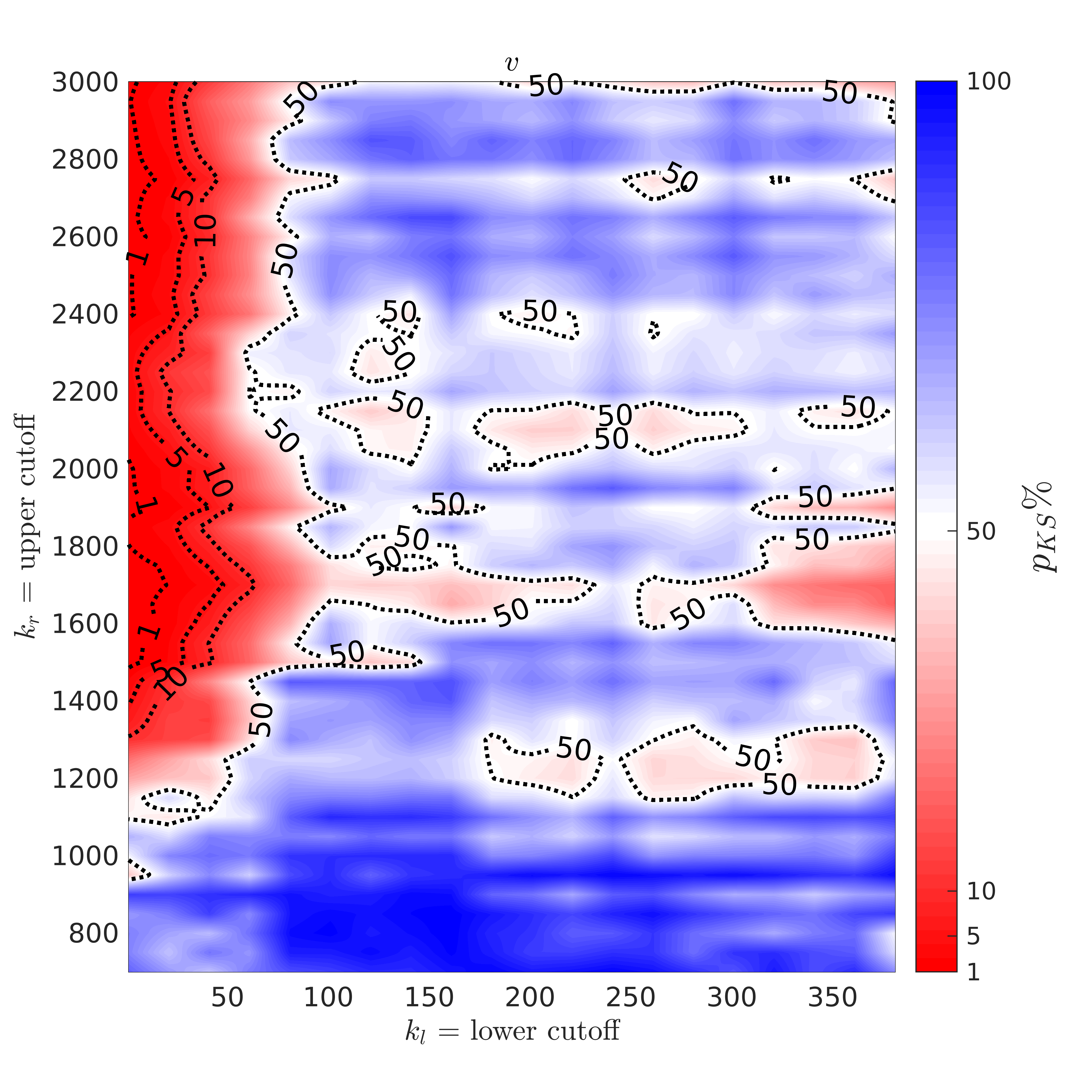}
    \caption{Dependence on the left and right window limits $k_l$ and $k_r$ of the KS-test $p_{KS}$-value in percent.}
    \label{fig:scan_v_pval}
  \end{figure}

Figure \ref{fig:scan_v_alpha} shows that the power-law exponent can vary between $-3$ and $0$ as a function of the left and right window limits $k_l$ and $k_r$. For a fixed upper limit $k_r$, the power-law exponent almost vanishes when the lower limit $k_l$ is low, which is explained by the fact that the periodogram flattens between $1-100$ on a log-log scale and only the exponential term may generate this behaviour. The power-law exponent decreases to around $-2.25$ as the left cutoff $k_l$ is raised to $k_l\approx 180$. Beyond this value, the evolution of $\alpha$ depends on the right cutoff $k_r$; $\alpha$ becomes even more negative if the right limit is below $1100$ or in the range of $1400-1700$, but stays constant if the mode number range is $2000-2400$.

Figure \ref{fig:scan_v_erra} illustrates the relative error in percent for the values of $\alpha$ - the exponent of the power-law in the compound function. The error of $\alpha$ varies from $4\%$ to $20\%$ when the left cut-off value $k_l$ varies from $100$ to $300$, and the right cut-off value $k_r$ varies from $1000$ to $3000$. The error of $\alpha$ is significant ($>20\%$) for $k_l<100$ and for $k_l>300$ and when the values of $k_r$ are relativelty low, $k_r\sim 800-1000$. The high error of $\alpha$ for $k_l<100$ and $k_r\sim 800-1000$ can be attributed to departures of the experimental spectrum from the power-law for small wave-vector values. The departures can be caused by, e.g., the effect of deterministric and initial conditions, and are indicative that for a short dynamics range with $[k_l,k_r]\sim[300,800]$, $\log_{10}(800/300)\approx 0.4$, the parameters of the compound function are a challenge to accurately identify. For $100<k_l<300$ and $1400<k_r<3000$, the error of $\alpha$ is insignificant, and is less than $10\%$. It overall decreases with the increase of $k_r$. This dependence of the error of $\alpha$ on $k_r$ suggests that in order to accurately ($<10\%$) identify the exponent $\alpha$ of the power-law component of the compound function, the statistical analysis should accound for the significant number of high frequency modes on the right-end ($k_r$) of the periodogram. 

On Figure \ref{fig:scan_v_beta}, the exponential decay rate is shown to vary as a function of the mode number window inversely to the power-law exponent, reaching values between $-7\times 10^{-3}$ and $0$. The $\alpha$ and $\beta$ estimates being positions of maximum likelihood, their variation is correlated with respect to changes in parameters such as $k_l$ and $k_r$. The correlation can be understood by considering the log of the power density spectrum $\ln S_{RT} = \alpha \ln k + \beta k + \gamma $ as the weighted sum of the three basic functions $\ln k$, $k$ and $1$, where the power-law has an influence on the exponential term and vice-versa under the projection method that is MLE.

Figure \ref{fig:scan_v_errb} illustrates the relative error in percent for the value of $\beta$ - the rate of the exponential decay in the compound function. The error of $\beta$ varies from $4\%$ to $10\%$, when the left cut-off value $k_l$ is less than $200$ and the right cut-off value $k_r$ varies from $800$ to $3000$. The error of $\beta$ varies from $10\%$ to $20\%$, when the left cut-off value $k_l$ varies from $200$ to $400$, and the right cut-off value $k_r$ varies from $2000$ to $3000$. The error of $\beta$ is significant, when the right cut-off values $k_r$ are relatively low, $800<kr<1600$, the left cut-off values $k_l$ are relatively high, $200<k_l<400$. This high error of $\beta$ indicates that for a short dynamic range with $[kl,kr]\sim[200,800]$, $\log_{10}(800/200)\approx 0.6$, the parameters of the compound function are a challenge to accurately identify. For $kl<200$ and $2000<kr<3000$, the error of $\beta$ is insignificant, and is less than $10\%$, and it overall decreases with the decrease of $k_l$. This dependence of the error of $\beta$ on $k_l$ suggests that in order to accurately ($<10\%$) identify the decay rate $\beta$ of the exponential component of the compound function the statistical analysis should account for the significant number of low frequency modes on the left end ($k_l$) of the periodogram.

Figure \ref{fig:scan_v_pval} displays the variation of the $p$-value of the KS test. The rejection region extends to a nearly rectangle bound by $k_l<80$ and $k_r>1300$. In the rejection region, with $0<k_l<80$ and $1300<k_l<3000$, the value of $p_{KS}$ sharply decreases from $50\%$ to $0\%$. This decrease can be attributed to the departures of the experimental spectrum from the compound function, which is caused by, e.g., the deterministic and initial conditions. The value of $p_{KS}$ varies from $50\%$ to $100\%$, when, overall, the left cut-off values are $k_l>80$ and the right cut-off values are $kr>1300$. This dependence of the $p_{KS}$-value on the left cut-off $k_l$ and right cut-off $k_r$ suggests that in order for the residuals (the deviations from the fitted spectrum) to be distributed according to the assumptions of the fitting technique, the statistical analysis should account for the significant number of high and low frequency modes on the left ($k_l$) and on the right ($k_r$) ends of the periodogram.

\begin{figure}
  \begin{subfigure}[t]{0.48\textwidth}
    \includegraphics[width=\columnwidth]{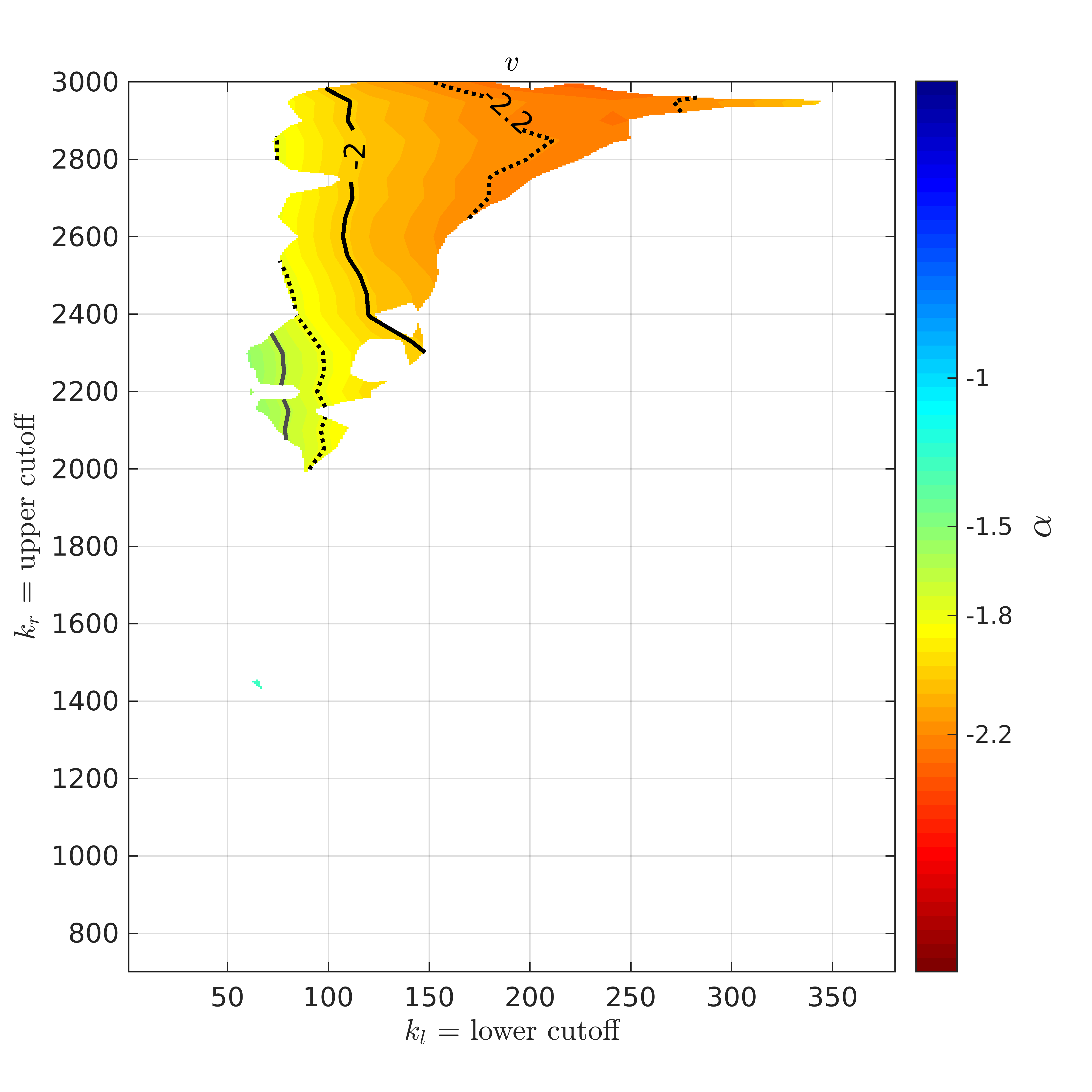}
    \caption{Power law exponent $\alpha$ in intersection of intervals with errors $<10\%$ and $p_{KS}>50\%$.}
    \label{fig:itsct_a}
  \end{subfigure}
  \begin{subfigure}[t]{0.48\textwidth}
    \includegraphics[width=\columnwidth]{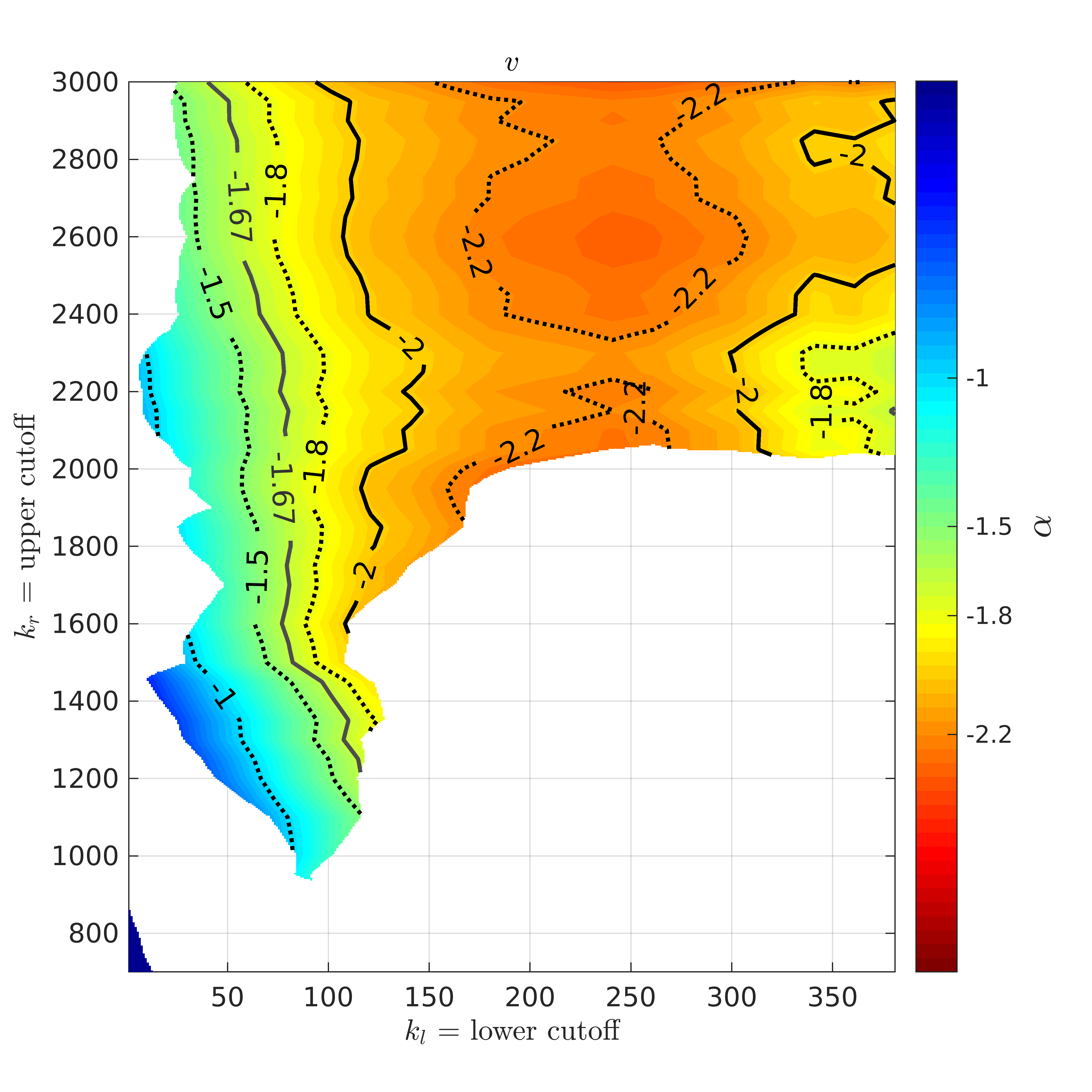}
    \caption{Power law exponent $\alpha$ in intersection of intervals with errors $<20\%$ and $p_{KS}>5\%$.}
    \label{fig:itsct2_a}
  \end{subfigure}
    \caption{Dependence on the left and right window limits $k_l$ and $k_r$ of the estimated fitting parameter $\alpha$.}
  \end{figure}

Figure \ref{fig:itsct_a} summarizes our investigation and presents the exponent alpha of the power-law
component of the compound function in the region, which is the intersection of the domains of $[k_l, k_r]$, where the error of $\alpha$ is less than $10\%$, the error of $\beta$ is less than $10\%$ and the $p_{KS}$-value is more than $50\%$. In this domain the compound function parameters $\alpha$ and $\beta$ are identified accurately, the goodness-of-fit is excellent, and the exponent $\alpha$ of the power-law is unambiguously defined over the dynamic range spanning $\log_{10}(3000/100)\approx 1.5$ decade, in consistency with the theory and with the results in Figure \ref{fig:period_v}. Note that for $\beta=\beta^*=1.00\times 10^{-3}$ with $1/\beta^*=1.00\times 10^3$, the corresponding dimensional wavevector $k^*=(\beta^* H)^{-1}=8.33\times 10^2[m^{-1}]$ and the length-scale $l^*=2\pi/k^*=7.43\times 10^{-3}[m]$ are comparable to the viscuous scales $k_\nu$ and $l_\nu$, $k^*\sim k_\nu$ and $l^*\sim l_\nu$, with $k_\nu/k^*=l^*/l_\nu=(1.35-1.29)$, in agreement with the theory.

Figure \ref{fig:itsct2_a} presents the exponent alpha of the power-law component of the compound function in a broader domain, which is the intersection of the intervals of $[k_l , k_r]$, where the error of $\alpha$ is less than $20\%$, the error of $\beta$ is less than $20\%$ and the $p_{KS}$-value is more than $5\%$. In this domain the exponent $\alpha$ is flexibly defined.

This work focuses on the fluctuations spectra of the $v$ component of the velocity. Similar analyses can be conducted for flucutations of other velocity components as well as for the the density field fluctuations. We leave the detailed investigation of statistical properties of the flow field in RT mixing for future work. Briefly, the fluctuations spectra of the velocity and the density can be described by the compound function $S(k)\sim k^\alpha \exp(\beta k)$. The values of the compound function parameters $\alpha$ and $\beta$ are demonstrably distrinct for fluctuations of various components of the velocity and the density and thus reveal anisotropy of RT mixing and its sensitivity to the deterministic conditions, in agreement with~\cite{abarzhi-2010a,abarzhi-2010b,meshkov-abarzhi-2019}.

\section{Properties of the power-law spectra}
\begin{figure}
  \begin{subfigure}[t]{0.48\textwidth}
    \includegraphics[width=\columnwidth]{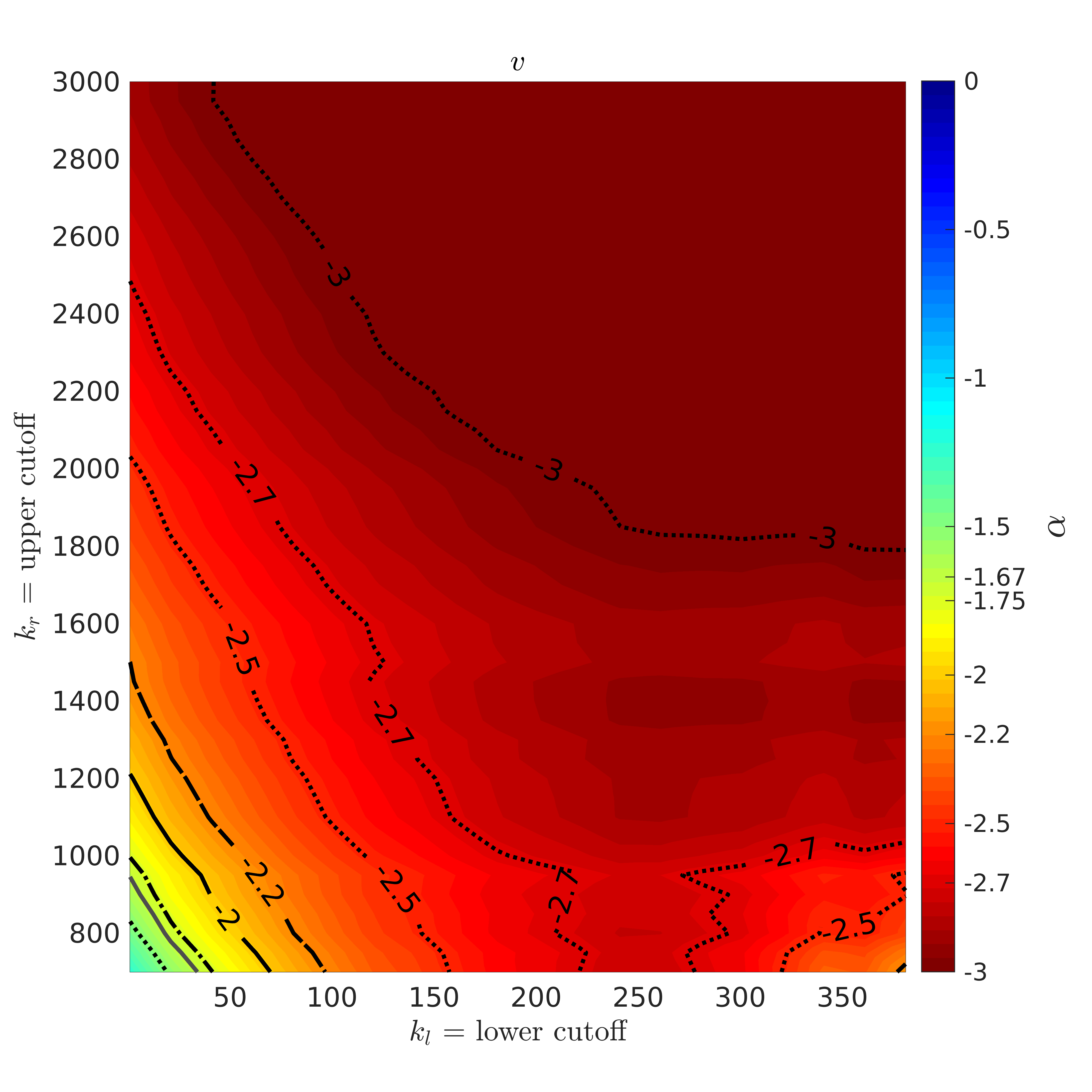}
    \caption{Power law exponent $\alpha$.}
    \label{fig:scan_v_alpha_betazero}
  \end{subfigure}
  \begin{subfigure}[t]{0.48\textwidth}
    \includegraphics[width=\columnwidth]{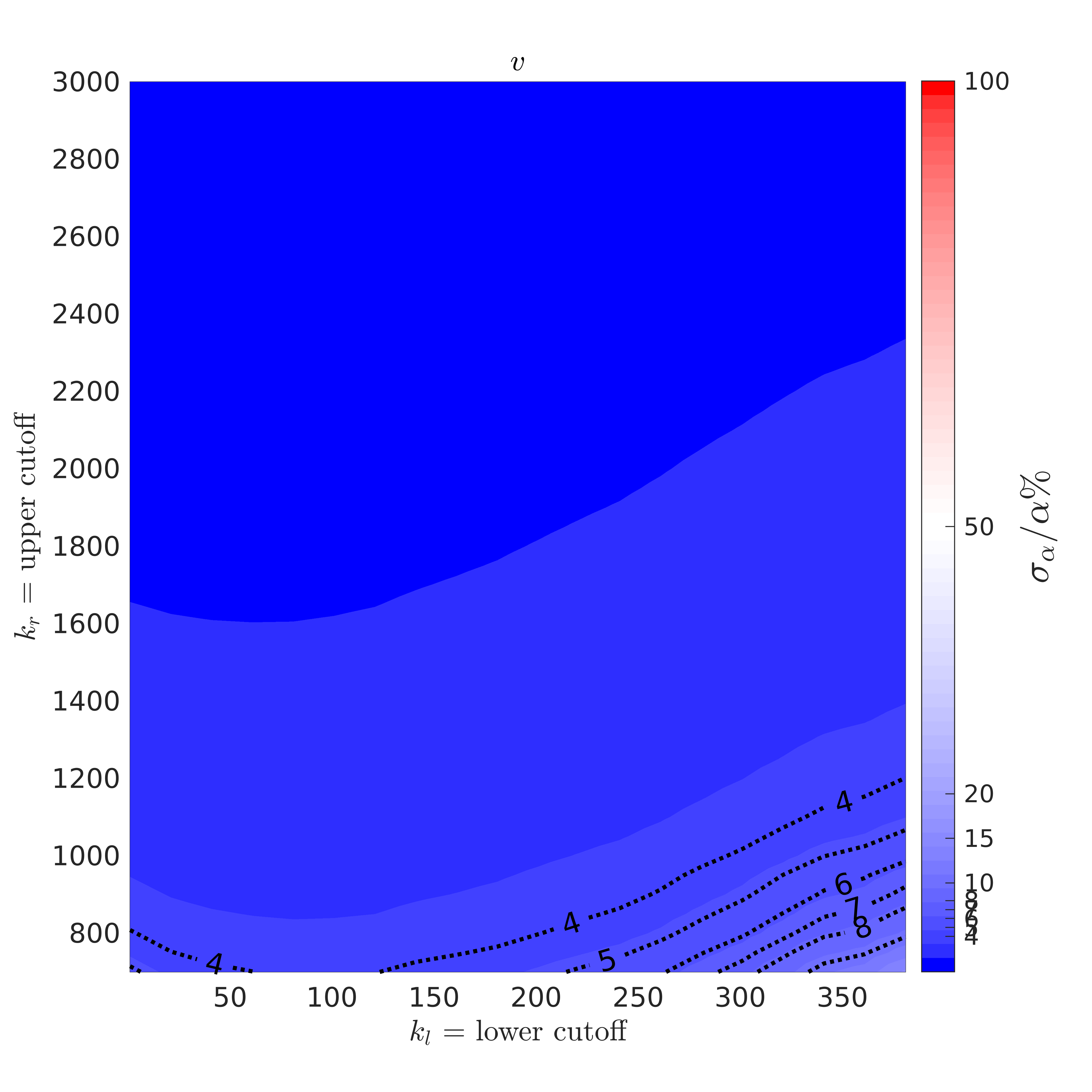}
    \caption{Relative error $\sigma_\alpha/\alpha$.}
    \label{fig:scan_v_erra_betazero}
  \end{subfigure}
  \caption{Dependence on the left and right window limits $k_l$ and $k_r$ of the estimated fiting parameters $\alpha$ and its relative error $\sigma_\alpha/\alpha$ in the MLE fits where the model spectrum is considered to be a power-law $S_{RT}(k)=C k^\alpha$ only.\label{fig:scan_v_betazero}}
\end{figure}

\begin{figure}
  \includegraphics[width=0.48\textwidth]{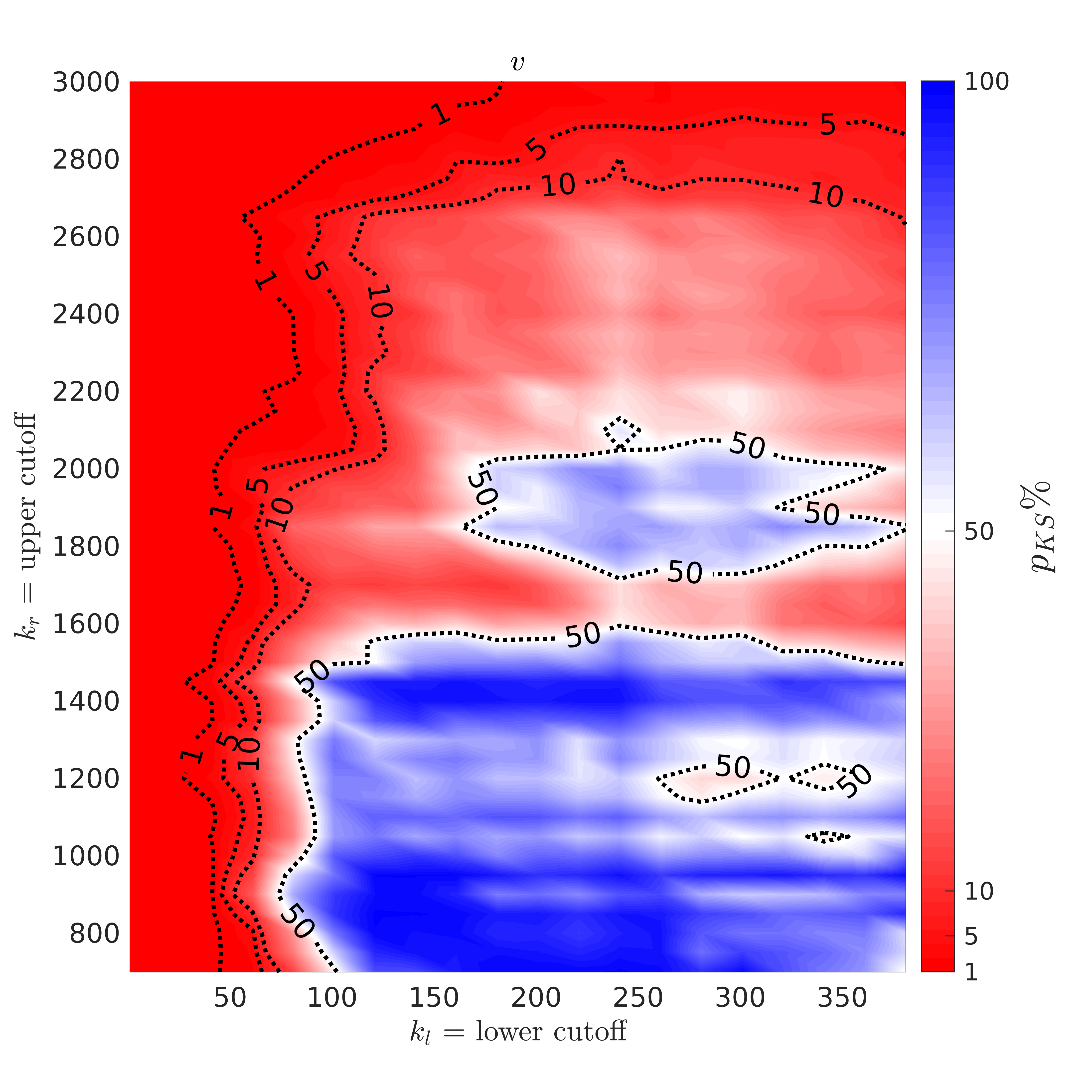}
    \caption{Dependence on the left and right window limits $k_l$ and $k_r$  on the KS-test $p_{KS}$-value in percent in the MLE fits where the model spectrum is considered to be a power-law $S_{RT}(k)=C k^\alpha$ only.}
    \label{fig:scan_v_pval_betazero}
\end{figure}

In order to illustrate the importance of physics-based statistical analysis in the data interpretation, we apply our method to analyse the power-law spectra $k^\alpha$ of the velocity fluctuations in RT mixing and compare our results with those from the visual inspection method in the experiments~\cite{akula-2017}.

\begin{figure}[h]
  \begin{subfigure}[t]{0.48\textwidth}
    \includegraphics[width=\columnwidth]{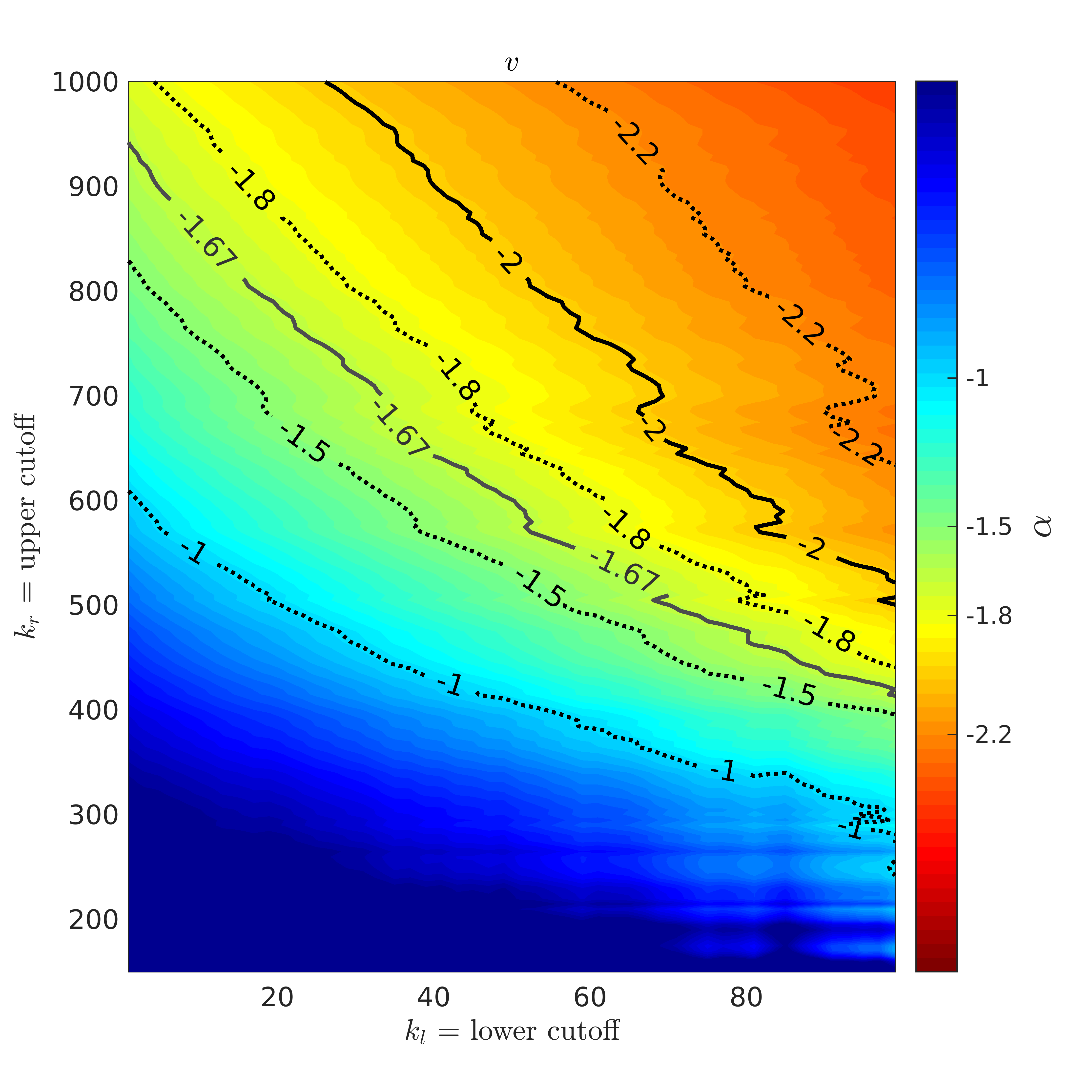}
    \caption{Power law exponent $\alpha$.}
    \label{fig:scan_v_alpha_betazerozoom}
  \end{subfigure}
  \begin{subfigure}[t]{0.48\textwidth}
    \includegraphics[width=\columnwidth]{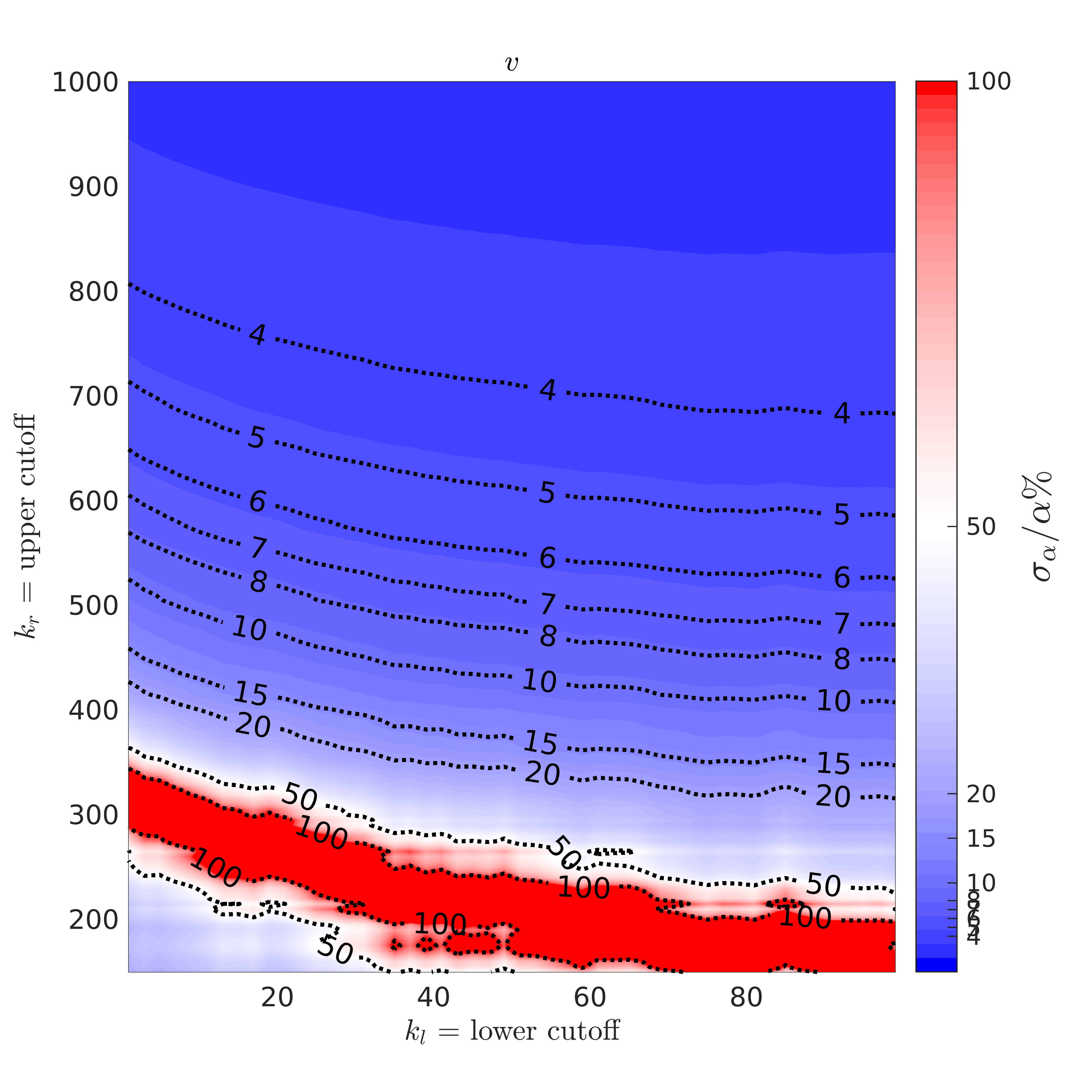}
    \caption{Relative error $\sigma/\alpha$ in percent.}
    \label{fig:scan_v_sigmaalpha_betazerozoom}
  \end{subfigure}
  \caption{Dependence on the left and right limits on a small ``sub-inertial'' range $k_l\in[1,100]$ and $k_r\in[150,1000]$ of the estimated fiting parameters $\alpha$ and its relative error $\sigma_\alpha/\alpha$ for the $v$-component of the velocity in the MLE fits where the model spectrum is considered to be a power-law $S_{RT}(k)=C k^\alpha$ only.\label{fig:scan_v_betazerozoom}}
\end{figure}

\begin{figure}
  \includegraphics[width=0.48\textwidth]{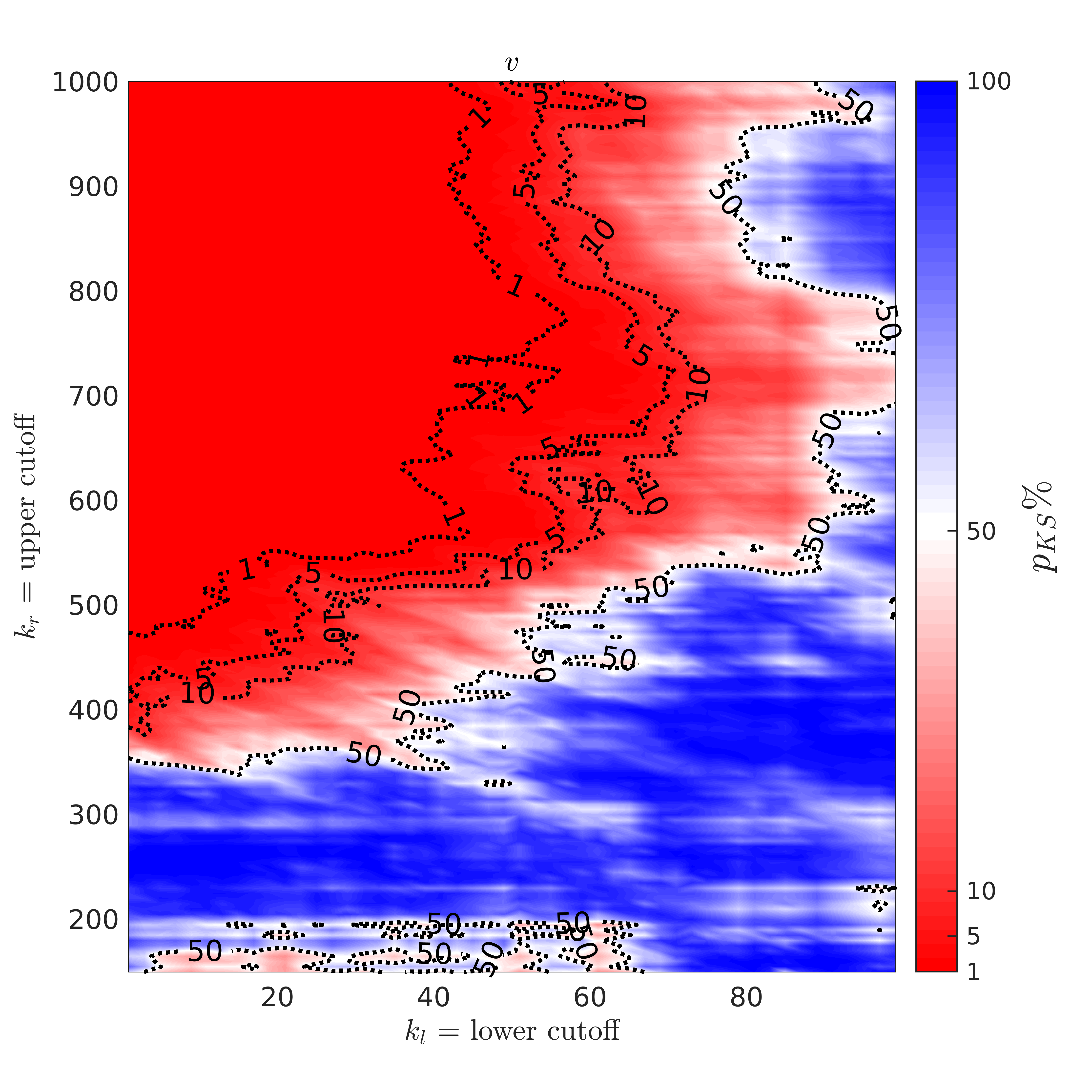}
  \caption{Dependence on the left and right limits on a small ``sub-inertial'' range $k_l\in[1,100]$ and $k_r\in[150,1000]$  of the KS-test $p_{KS}$-value for the $v$-component of the velocity in the MLE fits where the model spectrum is considered to be a power-law $S_{RT}(k)=C k^\alpha$ only.}
  \label{fig:scan_v_pval_betazerozoom}
\end{figure}

The experiments~\cite{akula-2017} processed the data by applying the MATLAB \code{smooth} function to the periodogram, used visual inspection to compare the fluctuations spectra with the scaling laws for various buoyant flows and for canonical turbulence, and concluded that in the 'pure' RT case the power-law spectra are steeper than $-5/3$. Figures \ref{fig:scan_v_alpha_betazero}, \ref{fig:scan_v_erra_betazero} and Figure \ref{fig:scan_v_pval_betazero} present our data analysis results for the power-law spectra $k^\alpha$ of the fluctuations of the $v$ component of the velocity in RT mixing. Figures \ref{fig:scan_v_alpha_betazero}, \ref{fig:scan_v_erra_betazero} and Figure \ref{fig:scan_v_pval_betazero} show respectively the values of $\alpha$ of the power-law exponent, the relative error of $\alpha$ in percent and the $p_{KS}$ goodness-of-fit value in percent in the fitting intervals $[k_l,k_r]$ for a broad range of delimiting values of the left cut-off $k_l$ and right cut-off $k_r$. Figures \ref{fig:scan_v_alpha_betazerozoom}, \ref{fig:scan_v_sigmaalpha_betazerozoom} and Figure \ref{fig:scan_v_pval_betazerozoom} show respectively the values of $\alpha$, the relative error of $\alpha$ and the $p_{KS}$-value, zoomed-in on the narrower intervals $[k_l, k_r]$, which are chosen in the experiments \cite{akula-2017} for the estimates of the power-law exponent.

%The green dot in Figures \ref{fig:scan_v_alpha_betazerozoom}, \ref{fig:scan_v_sigmaalpha_betazerozoom} and \ref{fig:scan_v_pval_betazerozoom} marks the interval $[k_l, k_r]$ selected by the experiments \cite{akula-2017} to illustrate compensated spectra.

%%

While these results are generally consistent with the conclusions~\cite{akula-2017}, the physics-based statistical data analysis approach identifies important properties of RT mixing that are challenging to see by the visual inspection method.

Similary to our results for the compound function $S(k)\sim k^\alpha\exp(\beta k)$ in Figure \ref{fig:period_v} and Figure \ref{fig:smo_v}, in case of the power-law function $S(k)\sim k^\alpha$ the MATLAB data smoothing has no influence on the value of the parameter $\alpha$. Yet, by making the signal 'crisper', it completely ruins the goodness-of-fit and makes the KS test reject the fitting every single time. We see that the analysis of unprocessed raw data is necessary to obtain statistically confident results, Figures \ref{fig:scan_v_betazero}, \ref{fig:scan_v_pval_betazero}, \ref{fig:scan_v_betazerozoom} and \ref{fig:scan_v_pval_betazerozoom}.

For the power-law fitting function $S(k)\sim k^\alpha$, our results further find that one needs to select very cautiously the range of values $[k_l,k_r]$ over which the fit is performed in order to obtain the scaling law $\sim k^{-5/3}$~\cite{kolmogorov-1941a,kolmogorov-1941b} and that this selected range spans less than a decade, in agreement with the conclusions of~\cite{akula-2017}. Power-law functions $\sim k^\alpha$ are scale-invariant and a free from characteristic scales naturally defining the borders of the fitting interval $[k_l,k_r]$. We see that for the data set with the short dynamics range, it is necessary to analyse the effect on the fitting function parameters of the range of $k$ values and the left and right cutoffs $k_{l(r)}$ in order to obtain reliable results, Figures \ref{fig:scan_v_betazero}, \ref{fig:scan_v_pval_betazero}, \ref{fig:scan_v_betazerozoom} and \ref{fig:scan_v_pval_betazerozoom}.

Our results in Figures \ref{fig:scan_v_betazero}, \ref{fig:scan_v_pval_betazero}, \ref{fig:scan_v_betazerozoom} and \ref{fig:scan_v_pval_betazerozoom} illustrate that for the power-law fitting function $\sim k^\alpha$ the fluctuation spectra are steeper than the canonical turbulence scaling law $\sim k^{-5/3}$. This result is consitent with the conclusions of~\cite{akula-2017}.

According to our results Figures \ref{fig:scan_v_betazero}, \ref{fig:scan_v_pval_betazero}, \ref{fig:scan_v_betazerozoom} and \ref{fig:scan_v_pval_betazerozoom} for the fluctuation of the $v$ component of the velocity in RT mixing, and for the power-law spectral function $\sim k^\alpha$, the statistically confident value of the power-law exponent departs from the values of the exponent $-5/3$ for the canonical turbulence~\cite{kolmogorov-1941a,kolmogorov-1941b,landau-lifshitz-1987}, the exponent $-7/4$ proposed for RT mixing by turbulent models~\cite{mikaelian-1989,zhou-2001}, the exponent $-2$ found by group theory for mathematical RT mixing~\cite{abarzhi-2010a,abarzhi-2010b}, the exponent $-11/5$ identified for atmospherical turbulence~\cite{obukhov-1959,bolgiano-1959}, and the exponent $-3$ obtained for homogenenous two-dimensional turbulence~\cite{batchelor-1969}.

By accounting for the properties of realistic fluids and the finite span of scales in $k$ in the experiments~\cite{akula-2017}, we apply the compound function $S(k)\sim k^\alpha\exp(\beta k)$ to explain the spectral properties of the fluctuations of the $v$ component of the velocity in RT mixing. The presence of the exponential terms $\exp(\beta k)$ and the power-law $k^\alpha$ with steeper than $-5/3$ exponent in the compound function $S(k)\sim k^\alpha\exp(\beta k)$ indicate that the dynamics of RT mixing, while self-similar, is more sensitive to the determinisitic (the initial and the flow) conditions, in agreement with~\cite{abarzhi-2010a,abarzhi-2010b,anisimov-2013,abarzhi-2019,meshkov-abarzhi-2019}.

\section{Discussion}
\label{sec:discussion} We have studied the spectral properties of Rayleigh-Taylor mixing by analysing experimental hot-wire anemometry data, Figures~\ref{fig:exp-timeseries}-\ref{fig:scan_v_pval_betazerozoom}. Guided by group theory, we have developed a formal statistical procedure for fitting the parameters of a given model power density spectrum to experimental time series, Eqs.~(\ref{eq:dft}-\ref{eq:reject-crit}). The method applies Maximum-Likelihood Estimation to evaluate the model parameters, the standard error and the goodness-of-fit. For the latter, the Kolmogorov-Smirnov test has been employed. The instrumental noise at the high-end of the spectrum has been incorporated into the model through a low level of white noise. The dependence of the fit parameters on the range of mode numbers, particularly on the left and right cutoffs, has been thoroughly investigated, including the values of the fit parameters, their relative errors and the goodness-of-fit. We have considered the sensitivity of the parameter estimations to the span of scales, the left and right cutoffs, and the choice of noise level, Figures~\ref{fig:exp-timeseries}-\ref{fig:scan_v_pval_betazerozoom}.

Bias-free methods of analysis and systematic interpretation of experimental and numerical data is necessary to ``get knowledge from the data'' of Rayleigh-Taylor mixing and to better understand Rayleigh-Taylor-relevant phenomena in nature and technology~\cite{abarzhi-2010a,abarzhi-2010b}. Our work is the first (to the authors' knowledge) to approach this task~\cite{abarzhi-2013}. Our analysis of hot-wire anemometry data finds that the power density spectrum of experimental quantities is described by the product of a power-law and an exponential, Figures \ref{fig:exp-timeseries}-\ref{fig:scan_v_pval_betazerozoom}. In the self-similar sub-range, Rayleigh-Taylor spectra are steeper than those of canonical turbulence, suggesting that RT mixing has stronger correlation and weaker fluctuations when compared to canonical turbulence. In the scale-dependent sub-range, the spectra are exponential rather than power-law, suggesting chaotic rather than stochastic behaviour of the fluctuations in Rayleigh-Taylor mixing. These results agree with group theory analysis~\cite{abarzhi-2010a,abarzhi-2010b,anisimov-2013,abarzhi-2005,meshkov-abarzhi-2019}. They are also consistent with the existence of anomalous scaling in realistic experimental spectra of canonical turbulence~\cite{sreenivasan-2018,sreenivasan-1999}.

Our analysis of the experimental hot-wire anemometry data has applied a number of assumptions, Eqs.~(\ref{eq:dft}-\ref{eq:reject-crit}). Particularly, the spectrum of fluctuations is an accurate diagnostics of statistically steady turbulence, whereas Rayleigh-Taylor mixing is statistically unsteady~\citep{abarzhi-2010a,abarzhi-2010b,anisimov-2013,sreenivasan-2018,abarzhi-2005,meshkov-abarzhi-2019}. Fluctuations in the wire resistance can be viewed as fluctuations of specific kinetic energy in canonical turbulence; a more accurate consideration may be required for Rayleigh-Taylor mixing with strongly changing scalar and vector fields~\citep{abarzhi-2010a,abarzhi-2010b,orlov-2010,meshkov-abarzhi-2019}. Maximum-likelihood estimations impose strong requirements on statistical properties of times series; these requirements may be challenging to obey in Rayleigh-Taylor mixing~\citep{abarzhi-2010a,abarzhi-2010b,abarzhi-2013,anisimov-2013,contreras-cristan-2006,brillinger-2001,whittle-1957,massey-1951,meshkov-abarzhi-2019}. The Kolmogorov-Smirnov test reliably quantifies goodness-of-fit in a multi-parameter system fluctuating about its mean; more caution may be required to quantify goodness-of-fit of fluctuations (in a sense - a noise of the noise)~\citep{abarzhi-2010a,abarzhi-2010b,abarzhi-2013,anisimov-2013,kolmogorov-1933,smirnov-1948}. Further developments are in demand on the fronts of experiment, theory, simulation and data analysis, in order to better understand the statistical properties of realistic non-equilibrium processes, such as anisotropic, inhomogeneous, statistically unsteady Rayleigh-Taylor mixing~\citep{abarzhi-2010a,abarzhi-2010b,abarzhi-2013,anisimov-2013,orlov-2010,sreenivasan-2018}.

Our work was focused on the development of a physics-based rigorous method of statistical analysis of raw data. Since Rayleigh-Taylor interfacial mixing is sensitive to the deterministic (the initial and the flow) conditions up to the Reynolds number $3.2\times 10^6$~\cite{abarzhi-2010a,abarzhi-2010b,meshkov-abarzhi-2019}, when analysing the velocity fluctuations data in the experiments~\cite{akula-2017}, we chose to study the pure Rayleigh-Taylor setup in order to separate the buoyancy effet from the shear; we analysed the data taken at late times in order to ensure that the flow is self-similar; we considered only one component of the velocity fluctuations, namely the component which is expected to be the least affected by the initial and the flow conditions. We found that the velocity fluctuations can be described by a compound function presented as a product of a power-law and an exponential. Our results revealed that for the accurate determination of the spectral properties of fluctuations, it is required to: (i) consider the raw unprocessed data; (ii) scrupulously investigate the range of the wavevector values and the left and right cut-offs; (iii) analyse the residuals and the goodness-of-fit. The experiments~\cite{akula-2017} were designed to study the unstably stratified shear flows at Reynolds numbers up to $3.4\times 10^4$. Our data analysis method can be applied to conduct the comparative study of the fluctuations of the velocity components and the density, to investigate the coupling of Rayleigh-Taylor and Kelvin-Helmholtz dynamics, and to analyse the time evolution of the fluctuations' spectra in these and in other experiments~\cite{akula-2017}. We leave these detailed studies for future work.

Note that the technique presented in this work may require alterations to treat filtered and processed signals, since it relies on \begin{enumerate*}[label=(\roman*)]
\item the canonical relation between the Fourier coefficients of a zero-mean stationary time series and its power spectral density;
\item the asymptotic independence of the signal's covariance matrix as a function of mode number Eqs.~(\ref{eq:dft}-\ref{eq:reject-crit}).
\end{enumerate*}

To conclude, we have developed a method of analysis of spectral properties of Rayleigh-Taylor mixing from raw experimental data, and have found that, in agreement with the theory, the power density spectrum of experimental quantities is described by the product of a power-law and an exponential. Our results indicate that rigorous physics-based statistical methods can help researcher to see beyond visual inspection, to achieve a bias-free interpretation of results, and to better understand
Rayleigh-Taylor dynamics and RT relevant phenomena in nature and technology.

\begin{acknowledgments} This work is supported by the University of Western Australia (AUS) and the National Science Foundation (USA). The experimental data were provided by Dr. Bhanesh Akula, Mr. Prasoon Suchandra, Mr. Mark Mikhaeil and Dr. Devesh Ranjan, whose contributions and valuable time are warmly appreciated.
\end{acknowledgments}

\nocite{*}
\bibliography{biblio}

\begin{thebibliography}{64}%
\makeatletter
\providecommand \@ifxundefined [1]{%
 \@ifx{#1\undefined}
}%
\providecommand \@ifnum [1]{%
 \ifnum #1\expandafter \@firstoftwo
 \else \expandafter \@secondoftwo
 \fi
}%
\providecommand \@ifx [1]{%
 \ifx #1\expandafter \@firstoftwo
 \else \expandafter \@secondoftwo
 \fi
}%
\providecommand \natexlab [1]{#1}%
\providecommand \enquote  [1]{``#1''}%
\providecommand \bibnamefont  [1]{#1}%
\providecommand \bibfnamefont [1]{#1}%
\providecommand \citenamefont [1]{#1}%
\providecommand \href@noop [0]{\@secondoftwo}%
\providecommand \href [0]{\begingroup \@sanitize@url \@href}%
\providecommand \@href[1]{\@@startlink{#1}\@@href}%
\providecommand \@@href[1]{\endgroup#1\@@endlink}%
\providecommand \@sanitize@url [0]{\catcode `\\12\catcode `\$12\catcode
  `\&12\catcode `\#12\catcode `\^12\catcode `\_12\catcode `\%12\relax}%
\providecommand \@@startlink[1]{}%
\providecommand \@@endlink[0]{}%
\providecommand \url  [0]{\begingroup\@sanitize@url \@url }%
\providecommand \@url [1]{\endgroup\@href {#1}{\urlprefix }}%
\providecommand \urlprefix  [0]{URL }%
\providecommand \Eprint [0]{\href }%
\providecommand \doibase [0]{http://dx.doi.org/}%
\providecommand \selectlanguage [0]{\@gobble}%
\providecommand \bibinfo  [0]{\@secondoftwo}%
\providecommand \bibfield  [0]{\@secondoftwo}%
\providecommand \translation [1]{[#1]}%
\providecommand \BibitemOpen [0]{}%
\providecommand \bibitemStop [0]{}%
\providecommand \bibitemNoStop [0]{.\EOS\space}%
\providecommand \EOS [0]{\spacefactor3000\relax}%
\providecommand \BibitemShut  [1]{\csname bibitem#1\endcsname}%
\let\auto@bib@innerbib\@empty
%</preamble>
\bibitem [{\citenamefont {Strutt~(3rd Baron~Rayleigh)}(1883)}]{rayleigh-1883}%
  \BibitemOpen
  \bibfield  {author} {\bibinfo {author} {\bibfnamefont {J.~W.}\ \bibnamefont
  {Strutt~(3rd Baron~Rayleigh)}},\ }\href {\doibase 10.1112/plms/s1-14.1.170}
  {\bibfield  {journal} {\bibinfo  {journal} {Proceedings of the London
  Mathematical Society}\ }\textbf {\bibinfo {volume} {s1-14}},\ \bibinfo
  {pages} {170} (\bibinfo {year} {1883})}\BibitemShut {NoStop}%
\bibitem [{\citenamefont {Davies}\ and\ \citenamefont
  {Taylor}(1950)}]{davies-taylor-1950}%
  \BibitemOpen
  \bibfield  {author} {\bibinfo {author} {\bibfnamefont {R.~M.}\ \bibnamefont
  {Davies}}\ and\ \bibinfo {author} {\bibfnamefont {G.~I.}\ \bibnamefont
  {Taylor}},\ }\href {\doibase 10.1098/rspa.1950.0023} {\bibfield  {journal}
  {\bibinfo  {journal} {Proceedings of the Royal Society of London. Series A.
  Mathematical and Physical Sciences}\ }\textbf {\bibinfo {volume} {200}},\
  \bibinfo {pages} {375} (\bibinfo {year} {1950})}\BibitemShut {NoStop}%
\bibitem [{\citenamefont {Abarzhi}(2010{\natexlab{a}})}]{abarzhi-2010a}%
  \BibitemOpen
  \bibfield  {author} {\bibinfo {author} {\bibfnamefont {S.~I.}\ \bibnamefont
  {Abarzhi}},\ }\href {\doibase 10.1098/rsta.2010.0020} {\bibfield  {journal}
  {\bibinfo  {journal} {Philosophical Transactions of the Royal Society A:
  Mathematical, Physical and Engineering Sciences}\ }\textbf {\bibinfo {volume}
  {368}},\ \bibinfo {pages} {1809} (\bibinfo {year}
  {2010}{\natexlab{a}})}\BibitemShut {NoStop}%
\bibitem [{\citenamefont {Abarzhi}(2010{\natexlab{b}})}]{abarzhi-2010b}%
  \BibitemOpen
  \bibfield  {author} {\bibinfo {author} {\bibfnamefont {S.~I.}\ \bibnamefont
  {Abarzhi}},\ }\href {\doibase 10.1209/0295-5075/91/35001} {\bibfield
  {journal} {\bibinfo  {journal} {{EPL} (Europhysics Letters)}\ }\textbf
  {\bibinfo {volume} {91}},\ \bibinfo {pages} {35001} (\bibinfo {year}
  {2010}{\natexlab{b}})}\BibitemShut {NoStop}%
\bibitem [{\citenamefont {Meshkov}(2006)}]{meshkov-2006}%
  \BibitemOpen
  \bibfield  {author} {\bibinfo {author} {\bibfnamefont {E.~E.}\ \bibnamefont
  {Meshkov}},\ }\href@noop {} {\emph {\bibinfo {title} {Studies of hydrodynamic
  instabilities in laboratory experiments}}}\ (\bibinfo  {publisher} {Sarov,
  FGUC-VNIIEF, ISBN 5-9515-0069-9, in Russian},\ \bibinfo {year}
  {2006})\BibitemShut {NoStop}%
\bibitem [{\citenamefont {Akula}\ \emph {et~al.}(2017)\citenamefont {Akula},
  \citenamefont {Suchandra}, \citenamefont {Mikhaeil},\ and\ \citenamefont
  {Ranjan}}]{akula-2017}%
  \BibitemOpen
  \bibfield  {author} {\bibinfo {author} {\bibfnamefont {B.}~\bibnamefont
  {Akula}}, \bibinfo {author} {\bibfnamefont {P.}~\bibnamefont {Suchandra}},
  \bibinfo {author} {\bibfnamefont {M.}~\bibnamefont {Mikhaeil}}, \ and\
  \bibinfo {author} {\bibfnamefont {D.}~\bibnamefont {Ranjan}},\ }\href
  {\doibase 10.1017/jfm.2017.95} {\bibfield  {journal} {\bibinfo  {journal}
  {Journal of Fluid Mechanics}\ }\textbf {\bibinfo {volume} {816}},\ \bibinfo
  {pages} {619–660} (\bibinfo {year} {2017})}\BibitemShut {NoStop}%
\bibitem [{\citenamefont {Akula}\ and\ \citenamefont
  {Ranjan}(2016)}]{akula-2016}%
  \BibitemOpen
  \bibfield  {author} {\bibinfo {author} {\bibfnamefont {B.}~\bibnamefont
  {Akula}}\ and\ \bibinfo {author} {\bibfnamefont {D.}~\bibnamefont {Ranjan}},\
  }\href {\doibase 10.1017/jfm.2016.199} {\bibfield  {journal} {\bibinfo
  {journal} {Journal of Fluid Mechanics}\ }\textbf {\bibinfo {volume} {795}},\
  \bibinfo {pages} {313–355} (\bibinfo {year} {2016})}\BibitemShut {NoStop}%
\bibitem [{\citenamefont {Abarzhi}\ \emph {et~al.}(2013)\citenamefont
  {Abarzhi}, \citenamefont {Gauthier},\ and\ \citenamefont
  {Sreenivasan}}]{abarzhi-2013}%
  \BibitemOpen
  \bibfield  {author} {\bibinfo {author} {\bibfnamefont {S.~I.}\ \bibnamefont
  {Abarzhi}}, \bibinfo {author} {\bibfnamefont {S.}~\bibnamefont {Gauthier}}, \
  and\ \bibinfo {author} {\bibfnamefont {K.~R.}\ \bibnamefont {Sreenivasan}},\
  }\href {\doibase 10.1098/rsta.2012.0436} {\bibfield  {journal} {\bibinfo
  {journal} {Philosophical Transactions of the Royal Society A: Mathematical,
  Physical and Engineering Sciences}\ }\textbf {\bibinfo {volume} {371}},\
  \bibinfo {pages} {20120436} (\bibinfo {year} {2013})}\BibitemShut {NoStop}%
\bibitem [{\citenamefont {Arnett}(1996)}]{arnett-1996}%
  \BibitemOpen
  \bibfield  {author} {\bibinfo {author} {\bibfnamefont {W.~D.}\ \bibnamefont
  {Arnett}},\ }\href@noop {} {\emph {\bibinfo {title} {Supernovae and
  Nucleosynthesis: An Investigation of the History of Matter, from the Big Bang
  to the Present}}},\ Princeton Series in Astrophysics\ (\bibinfo  {publisher}
  {Princeton University Press},\ \bibinfo {year} {1996})\BibitemShut {NoStop}%
\bibitem [{\citenamefont {Haan}\ \emph {et~al.}(2011)\citenamefont {Haan},
  \citenamefont {Lindl}, \citenamefont {Callahan}, \citenamefont {Clark},
  \citenamefont {Salmonson}, \citenamefont {Hammel}, \citenamefont {Atherton},
  \citenamefont {Cook}, \citenamefont {Edwards}, \citenamefont {Glenzer},
  \citenamefont {Hamza}, \citenamefont {Hatchett}, \citenamefont {Herrmann},
  \citenamefont {Hinkel}, \citenamefont {Ho}, \citenamefont {Huang},
  \citenamefont {Jones}, \citenamefont {Kline}, \citenamefont {Kyrala},
  \citenamefont {Landen}, \citenamefont {MacGowan}, \citenamefont {Marinak},
  \citenamefont {Meyerhofer}, \citenamefont {Milovich}, \citenamefont {Moreno},
  \citenamefont {Moses}, \citenamefont {Munro}, \citenamefont {Nikroo},
  \citenamefont {Olson}, \citenamefont {Peterson}, \citenamefont {Pollaine},
  \citenamefont {Ralph}, \citenamefont {Robey}, \citenamefont {Spears},
  \citenamefont {Springer}, \citenamefont {Suter}, \citenamefont {Thomas},
  \citenamefont {Town}, \citenamefont {Vesey}, \citenamefont {Weber},
  \citenamefont {Wilkens},\ and\ \citenamefont {Wilson}}]{haan-2011}%
  \BibitemOpen
  \bibfield  {author} {\bibinfo {author} {\bibfnamefont {S.~W.}\ \bibnamefont
  {Haan}}, \bibinfo {author} {\bibfnamefont {J.~D.}\ \bibnamefont {Lindl}},
  \bibinfo {author} {\bibfnamefont {D.~A.}\ \bibnamefont {Callahan}}, \bibinfo
  {author} {\bibfnamefont {D.~S.}\ \bibnamefont {Clark}}, \bibinfo {author}
  {\bibfnamefont {J.~D.}\ \bibnamefont {Salmonson}}, \bibinfo {author}
  {\bibfnamefont {B.~A.}\ \bibnamefont {Hammel}}, \bibinfo {author}
  {\bibfnamefont {L.~J.}\ \bibnamefont {Atherton}}, \bibinfo {author}
  {\bibfnamefont {R.~C.}\ \bibnamefont {Cook}}, \bibinfo {author}
  {\bibfnamefont {M.~J.}\ \bibnamefont {Edwards}}, \bibinfo {author}
  {\bibfnamefont {S.}~\bibnamefont {Glenzer}}, \bibinfo {author} {\bibfnamefont
  {A.~V.}\ \bibnamefont {Hamza}}, \bibinfo {author} {\bibfnamefont {S.~P.}\
  \bibnamefont {Hatchett}}, \bibinfo {author} {\bibfnamefont {M.~C.}\
  \bibnamefont {Herrmann}}, \bibinfo {author} {\bibfnamefont {D.~E.}\
  \bibnamefont {Hinkel}}, \bibinfo {author} {\bibfnamefont {D.~D.}\
  \bibnamefont {Ho}}, \bibinfo {author} {\bibfnamefont {H.}~\bibnamefont
  {Huang}}, \bibinfo {author} {\bibfnamefont {O.~S.}\ \bibnamefont {Jones}},
  \bibinfo {author} {\bibfnamefont {J.}~\bibnamefont {Kline}}, \bibinfo
  {author} {\bibfnamefont {G.}~\bibnamefont {Kyrala}}, \bibinfo {author}
  {\bibfnamefont {O.~L.}\ \bibnamefont {Landen}}, \bibinfo {author}
  {\bibfnamefont {B.~J.}\ \bibnamefont {MacGowan}}, \bibinfo {author}
  {\bibfnamefont {M.~M.}\ \bibnamefont {Marinak}}, \bibinfo {author}
  {\bibfnamefont {D.~D.}\ \bibnamefont {Meyerhofer}}, \bibinfo {author}
  {\bibfnamefont {J.~L.}\ \bibnamefont {Milovich}}, \bibinfo {author}
  {\bibfnamefont {K.~A.}\ \bibnamefont {Moreno}}, \bibinfo {author}
  {\bibfnamefont {E.~I.}\ \bibnamefont {Moses}}, \bibinfo {author}
  {\bibfnamefont {D.~H.}\ \bibnamefont {Munro}}, \bibinfo {author}
  {\bibfnamefont {A.}~\bibnamefont {Nikroo}}, \bibinfo {author} {\bibfnamefont
  {R.~E.}\ \bibnamefont {Olson}}, \bibinfo {author} {\bibfnamefont
  {K.}~\bibnamefont {Peterson}}, \bibinfo {author} {\bibfnamefont {S.~M.}\
  \bibnamefont {Pollaine}}, \bibinfo {author} {\bibfnamefont {J.~E.}\
  \bibnamefont {Ralph}}, \bibinfo {author} {\bibfnamefont {H.~F.}\ \bibnamefont
  {Robey}}, \bibinfo {author} {\bibfnamefont {B.~K.}\ \bibnamefont {Spears}},
  \bibinfo {author} {\bibfnamefont {P.~T.}\ \bibnamefont {Springer}}, \bibinfo
  {author} {\bibfnamefont {L.~J.}\ \bibnamefont {Suter}}, \bibinfo {author}
  {\bibfnamefont {C.~A.}\ \bibnamefont {Thomas}}, \bibinfo {author}
  {\bibfnamefont {R.~P.}\ \bibnamefont {Town}}, \bibinfo {author}
  {\bibfnamefont {R.}~\bibnamefont {Vesey}}, \bibinfo {author} {\bibfnamefont
  {S.~V.}\ \bibnamefont {Weber}}, \bibinfo {author} {\bibfnamefont {H.~L.}\
  \bibnamefont {Wilkens}}, \ and\ \bibinfo {author} {\bibfnamefont {D.~C.}\
  \bibnamefont {Wilson}},\ }\href {\doibase 10.1063/1.3592169} {\bibfield
  {journal} {\bibinfo  {journal} {Physics of Plasmas}\ }\textbf {\bibinfo
  {volume} {18}},\ \bibinfo {pages} {051001} (\bibinfo {year}
  {2011})}\BibitemShut {NoStop}%
\bibitem [{\citenamefont {Peters}(2000)}]{peters-2000}%
  \BibitemOpen
  \bibfield  {author} {\bibinfo {author} {\bibfnamefont {N.}~\bibnamefont
  {Peters}},\ }\href {\doibase 10.1017/CBO9780511612701} {\emph {\bibinfo
  {title} {Turbulent Combustion}}},\ Cambridge Monographs on Mechanics\
  (\bibinfo  {publisher} {Cambridge University Press},\ \bibinfo {year}
  {2000})\BibitemShut {NoStop}%
\bibitem [{\citenamefont {Anisimov}\ \emph {et~al.}(2013)\citenamefont
  {Anisimov}, \citenamefont {Drake}, \citenamefont {Gauthier}, \citenamefont
  {Meshkov},\ and\ \citenamefont {Abarzhi}}]{anisimov-2013}%
  \BibitemOpen
  \bibfield  {author} {\bibinfo {author} {\bibfnamefont {S.~I.}\ \bibnamefont
  {Anisimov}}, \bibinfo {author} {\bibfnamefont {R.~P.}\ \bibnamefont {Drake}},
  \bibinfo {author} {\bibfnamefont {S.}~\bibnamefont {Gauthier}}, \bibinfo
  {author} {\bibfnamefont {E.~E.}\ \bibnamefont {Meshkov}}, \ and\ \bibinfo
  {author} {\bibfnamefont {S.~I.}\ \bibnamefont {Abarzhi}},\ }\href {\doibase
  10.1098/rsta.2013.0266} {\bibfield  {journal} {\bibinfo  {journal}
  {Philosophical Transactions of the Royal Society A: Mathematical, Physical
  and Engineering Sciences}\ }\textbf {\bibinfo {volume} {371}},\ \bibinfo
  {pages} {20130266} (\bibinfo {year} {2013})}\BibitemShut {NoStop}%
\bibitem [{\citenamefont {Orlov}\ \emph {et~al.}(2010)\citenamefont {Orlov},
  \citenamefont {Abarzhi}, \citenamefont {Oh}, \citenamefont {Barbastathis},\
  and\ \citenamefont {Sreenivasan}}]{orlov-2010}%
  \BibitemOpen
  \bibfield  {author} {\bibinfo {author} {\bibfnamefont {S.~S.}\ \bibnamefont
  {Orlov}}, \bibinfo {author} {\bibfnamefont {S.~I.}\ \bibnamefont {Abarzhi}},
  \bibinfo {author} {\bibfnamefont {S.~B.}\ \bibnamefont {Oh}}, \bibinfo
  {author} {\bibfnamefont {G.}~\bibnamefont {Barbastathis}}, \ and\ \bibinfo
  {author} {\bibfnamefont {K.~R.}\ \bibnamefont {Sreenivasan}},\ }\href
  {\doibase 10.1098/rsta.2009.0285} {\bibfield  {journal} {\bibinfo  {journal}
  {Philosophical Transactions of the Royal Society A: Mathematical, Physical
  and Engineering Sciences}\ }\textbf {\bibinfo {volume} {368}},\ \bibinfo
  {pages} {1705} (\bibinfo {year} {2010})}\BibitemShut {NoStop}%
\bibitem [{\citenamefont {Sreenivasan}(2018)}]{sreenivasan-2018}%
  \BibitemOpen
  \bibfield  {author} {\bibinfo {author} {\bibfnamefont {K.~R.}\ \bibnamefont
  {Sreenivasan}},\ }\href {\doibase 10.1073/pnas.1800463115} {\bibfield
  {journal} {\bibinfo  {journal} {Proceedings of the National Academy of
  Sciences of the USA}\ } (\bibinfo {year} {2018}),\
  10.1073/pnas.1800463115}\BibitemShut {NoStop}%
\bibitem [{\citenamefont {Meshkov}(2013)}]{meshkov-2013}%
  \BibitemOpen
  \bibfield  {author} {\bibinfo {author} {\bibfnamefont {E.~E.}\ \bibnamefont
  {Meshkov}},\ }\href {\doibase 10.1098/rsta.2012.0288} {\bibfield  {journal}
  {\bibinfo  {journal} {Philosophical Transactions of the Royal Society A:
  Mathematical, Physical and Engineering Sciences}\ }\textbf {\bibinfo {volume}
  {371}},\ \bibinfo {pages} {20120288} (\bibinfo {year} {2013})}\BibitemShut
  {NoStop}%
\bibitem [{\citenamefont {Robey}\ \emph {et~al.}(2003)\citenamefont {Robey},
  \citenamefont {Zhou}, \citenamefont {Buckingham}, \citenamefont {Keiter},
  \citenamefont {Remington},\ and\ \citenamefont {Drake}}]{robey-2003}%
  \BibitemOpen
  \bibfield  {author} {\bibinfo {author} {\bibfnamefont {H.~F.}\ \bibnamefont
  {Robey}}, \bibinfo {author} {\bibfnamefont {Y.}~\bibnamefont {Zhou}},
  \bibinfo {author} {\bibfnamefont {A.~C.}\ \bibnamefont {Buckingham}},
  \bibinfo {author} {\bibfnamefont {P.}~\bibnamefont {Keiter}}, \bibinfo
  {author} {\bibfnamefont {B.~A.}\ \bibnamefont {Remington}}, \ and\ \bibinfo
  {author} {\bibfnamefont {R.~P.}\ \bibnamefont {Drake}},\ }\href {\doibase
  10.1063/1.1534584} {\bibfield  {journal} {\bibinfo  {journal} {Physics of
  Plasmas}\ }\textbf {\bibinfo {volume} {10}},\ \bibinfo {pages} {614}
  (\bibinfo {year} {2003})}\BibitemShut {NoStop}%
\bibitem [{\citenamefont {Remington}\ \emph {et~al.}(2018)\citenamefont
  {Remington}, \citenamefont {Park}, \citenamefont {Casey}, \citenamefont
  {Cavallo}, \citenamefont {Clark}, \citenamefont {Huntington}, \citenamefont
  {Kuranz}, \citenamefont {Miles}, \citenamefont {Nagel}, \citenamefont
  {Raman},\ and\ \citenamefont {Smalyuk}}]{remington-2018}%
  \BibitemOpen
  \bibfield  {author} {\bibinfo {author} {\bibfnamefont {B.~A.}\ \bibnamefont
  {Remington}}, \bibinfo {author} {\bibfnamefont {H.-S.}\ \bibnamefont {Park}},
  \bibinfo {author} {\bibfnamefont {D.~T.}\ \bibnamefont {Casey}}, \bibinfo
  {author} {\bibfnamefont {R.~M.}\ \bibnamefont {Cavallo}}, \bibinfo {author}
  {\bibfnamefont {D.~S.}\ \bibnamefont {Clark}}, \bibinfo {author}
  {\bibfnamefont {C.~M.}\ \bibnamefont {Huntington}}, \bibinfo {author}
  {\bibfnamefont {C.~C.}\ \bibnamefont {Kuranz}}, \bibinfo {author}
  {\bibfnamefont {A.~R.}\ \bibnamefont {Miles}}, \bibinfo {author}
  {\bibfnamefont {S.~R.}\ \bibnamefont {Nagel}}, \bibinfo {author}
  {\bibfnamefont {K.~S.}\ \bibnamefont {Raman}}, \ and\ \bibinfo {author}
  {\bibfnamefont {V.~A.}\ \bibnamefont {Smalyuk}},\ }\href {\doibase
  10.1073/pnas.1717236115} {\bibfield  {journal} {\bibinfo  {journal}
  {Proceedings of the National Academy of Sciences of the USA}\ } (\bibinfo
  {year} {2018}),\ 10.1073/pnas.1717236115}\BibitemShut {NoStop}%
\bibitem [{\citenamefont {Ristorcelli}\ and\ \citenamefont
  {Clark}(2004)}]{ristorcelli-2004}%
  \BibitemOpen
  \bibfield  {author} {\bibinfo {author} {\bibfnamefont {J.~R.}\ \bibnamefont
  {Ristorcelli}}\ and\ \bibinfo {author} {\bibfnamefont {T.~T.}\ \bibnamefont
  {Clark}},\ }\href {\doibase 10.1017/S0022112004008286} {\bibfield  {journal}
  {\bibinfo  {journal} {Journal of Fluid Mechanics}\ }\textbf {\bibinfo
  {volume} {507}},\ \bibinfo {pages} {213–253} (\bibinfo {year}
  {2004})}\BibitemShut {NoStop}%
\bibitem [{\citenamefont {Glimm}\ \emph {et~al.}(2013)\citenamefont {Glimm},
  \citenamefont {Sharp}, \citenamefont {Kaman},\ and\ \citenamefont
  {Lim}}]{glimm-2013}%
  \BibitemOpen
  \bibfield  {author} {\bibinfo {author} {\bibfnamefont {J.}~\bibnamefont
  {Glimm}}, \bibinfo {author} {\bibfnamefont {D.~H.}\ \bibnamefont {Sharp}},
  \bibinfo {author} {\bibfnamefont {T.}~\bibnamefont {Kaman}}, \ and\ \bibinfo
  {author} {\bibfnamefont {H.}~\bibnamefont {Lim}},\ }\href {\doibase
  10.1098/rsta.2012.0183} {\bibfield  {journal} {\bibinfo  {journal}
  {Philosophical Transactions of the Royal Society A: Mathematical, Physical
  and Engineering Sciences}\ }\textbf {\bibinfo {volume} {371}},\ \bibinfo
  {pages} {20120183} (\bibinfo {year} {2013})}\BibitemShut {NoStop}%
\bibitem [{\citenamefont {Kadau}\ \emph {et~al.}(2010)\citenamefont {Kadau},
  \citenamefont {Barber}, \citenamefont {Germann}, \citenamefont {Holian},\
  and\ \citenamefont {Alder}}]{kadau-2010}%
  \BibitemOpen
  \bibfield  {author} {\bibinfo {author} {\bibfnamefont {K.}~\bibnamefont
  {Kadau}}, \bibinfo {author} {\bibfnamefont {J.~L.}\ \bibnamefont {Barber}},
  \bibinfo {author} {\bibfnamefont {T.~C.}\ \bibnamefont {Germann}}, \bibinfo
  {author} {\bibfnamefont {B.~L.}\ \bibnamefont {Holian}}, \ and\ \bibinfo
  {author} {\bibfnamefont {B.~J.}\ \bibnamefont {Alder}},\ }\href {\doibase
  10.1098/rsta.2009.0218} {\bibfield  {journal} {\bibinfo  {journal}
  {Philosophical Transactions of the Royal Society A: Mathematical, Physical
  and Engineering Sciences}\ }\textbf {\bibinfo {volume} {368}},\ \bibinfo
  {pages} {1547} (\bibinfo {year} {2010})}\BibitemShut {NoStop}%
\bibitem [{\citenamefont {Youngs}(2013)}]{youngs-2013}%
  \BibitemOpen
  \bibfield  {author} {\bibinfo {author} {\bibfnamefont {D.~L.}\ \bibnamefont
  {Youngs}},\ }\href {\doibase 10.1098/rsta.2012.0173} {\bibfield  {journal}
  {\bibinfo  {journal} {Philosophical Transactions of the Royal Society A:
  Mathematical, Physical and Engineering Sciences}\ }\textbf {\bibinfo {volume}
  {371}},\ \bibinfo {pages} {20120173} (\bibinfo {year} {2013})}\BibitemShut
  {NoStop}%
\bibitem [{\citenamefont {Abarzhi}\ \emph {et~al.}(2005)\citenamefont
  {Abarzhi}, \citenamefont {Gorobets},\ and\ \citenamefont
  {Sreenivasan}}]{abarzhi-2005}%
  \BibitemOpen
  \bibfield  {author} {\bibinfo {author} {\bibfnamefont {S.~I.}\ \bibnamefont
  {Abarzhi}}, \bibinfo {author} {\bibfnamefont {A.}~\bibnamefont {Gorobets}}, \
  and\ \bibinfo {author} {\bibfnamefont {K.~R.}\ \bibnamefont {Sreenivasan}},\
  }\href {\doibase 10.1063/1.2009027} {\bibfield  {journal} {\bibinfo
  {journal} {Physics of Fluids}\ }\textbf {\bibinfo {volume} {17}},\ \bibinfo
  {pages} {081705} (\bibinfo {year} {2005})}\BibitemShut {NoStop}%
\bibitem [{\citenamefont {Meshkov}\ and\ \citenamefont
  {Abarzhi}(2019)}]{meshkov-abarzhi-2019}%
  \BibitemOpen
  \bibfield  {author} {\bibinfo {author} {\bibfnamefont {E.~E.}\ \bibnamefont
  {Meshkov}}\ and\ \bibinfo {author} {\bibfnamefont {S.~I.}\ \bibnamefont
  {Abarzhi}},\ }\href {\doibase 10.1088/1873-7005/ab3e83} {\bibfield  {journal}
  {\bibinfo  {journal} {Fluid Dynamics Research}\ }\textbf {\bibinfo {volume}
  {51}},\ \bibinfo {pages} {065502} (\bibinfo {year} {2019})}\BibitemShut
  {NoStop}%
\bibitem [{\citenamefont {Abarzhi}\ \emph {et~al.}(2019)\citenamefont
  {Abarzhi}, \citenamefont {Bhowmick}, \citenamefont {Naveh}, \citenamefont
  {Pandian}, \citenamefont {Swisher}, \citenamefont {Stellingwerf},\ and\
  \citenamefont {Arnett}}]{abarzhi-2019}%
  \BibitemOpen
  \bibfield  {author} {\bibinfo {author} {\bibfnamefont {S.~I.}\ \bibnamefont
  {Abarzhi}}, \bibinfo {author} {\bibfnamefont {A.~K.}\ \bibnamefont
  {Bhowmick}}, \bibinfo {author} {\bibfnamefont {A.}~\bibnamefont {Naveh}},
  \bibinfo {author} {\bibfnamefont {A.}~\bibnamefont {Pandian}}, \bibinfo
  {author} {\bibfnamefont {N.~C.}\ \bibnamefont {Swisher}}, \bibinfo {author}
  {\bibfnamefont {R.~F.}\ \bibnamefont {Stellingwerf}}, \ and\ \bibinfo
  {author} {\bibfnamefont {W.~D.}\ \bibnamefont {Arnett}},\ }\href {\doibase
  10.1073/pnas.1714502115} {\bibfield  {journal} {\bibinfo  {journal}
  {Proceedings of the National Academy of Sciences}\ }\textbf {\bibinfo
  {volume} {116}},\ \bibinfo {pages} {18184} (\bibinfo {year}
  {2019})}\BibitemShut {NoStop}%
\bibitem [{\citenamefont {Kolmogorov}(1933)}]{kolmogorov-1933}%
  \BibitemOpen
  \bibfield  {author} {\bibinfo {author} {\bibfnamefont {A.~N.}\ \bibnamefont
  {Kolmogorov}},\ }\href@noop {} {\bibfield  {journal} {\bibinfo  {journal}
  {Giornale dell'Istituto Italiano degli Attuari}\ }\textbf {\bibinfo {volume}
  {4}},\ \bibinfo {pages} {83} (\bibinfo {year} {1933})}\BibitemShut {NoStop}%
\bibitem [{\citenamefont {Smirnov}(1948)}]{smirnov-1948}%
  \BibitemOpen
  \bibfield  {author} {\bibinfo {author} {\bibfnamefont {N.}~\bibnamefont
  {Smirnov}},\ }\href {\doibase 10.1214/aoms/1177730256} {\bibfield  {journal}
  {\bibinfo  {journal} {Ann. Math. Statist.}\ }\textbf {\bibinfo {volume}
  {19}},\ \bibinfo {pages} {279} (\bibinfo {year} {1948})}\BibitemShut
  {NoStop}%
\bibitem [{\citenamefont
  {{Kolmogorov}}(1941{\natexlab{a}})}]{kolmogorov-1941a}%
  \BibitemOpen
  \bibfield  {author} {\bibinfo {author} {\bibfnamefont {A.}~\bibnamefont
  {{Kolmogorov}}},\ }\href@noop {} {\bibfield  {journal} {\bibinfo  {journal}
  {Dokl. Akad. Nauk SSSR}\ }\textbf {\bibinfo {volume} {30}},\ \bibinfo {pages}
  {299} (\bibinfo {year} {1941}{\natexlab{a}})}\BibitemShut {NoStop}%
\bibitem [{\citenamefont
  {{Kolmogorov}}(1941{\natexlab{b}})}]{kolmogorov-1941b}%
  \BibitemOpen
  \bibfield  {author} {\bibinfo {author} {\bibfnamefont {A.}~\bibnamefont
  {{Kolmogorov}}},\ }\href@noop {} {\bibfield  {journal} {\bibinfo  {journal}
  {Dokl. Akad. Nauk SSSR}\ }\textbf {\bibinfo {volume} {31}},\ \bibinfo {pages}
  {538} (\bibinfo {year} {1941}{\natexlab{b}})}\BibitemShut {NoStop}%
\bibitem [{\citenamefont {Landau}\ and\ \citenamefont
  {Lifshitz}(1987)}]{landau-lifshitz-1987}%
  \BibitemOpen
  \bibfield  {author} {\bibinfo {author} {\bibfnamefont {L.}~\bibnamefont
  {Landau}}\ and\ \bibinfo {author} {\bibfnamefont {E.}~\bibnamefont
  {Lifshitz}},\ }\href@noop {} {\emph {\bibinfo {title} {Course of theoretical
  physics}}},\ Vol.\ \bibinfo {volume} {I-X}\ (\bibinfo  {publisher} {Elsevier
  Science},\ \bibinfo {year} {1987})\BibitemShut {NoStop}%
\bibitem [{\citenamefont {Sreenivasan}(1999)}]{sreenivasan-1999}%
  \BibitemOpen
  \bibfield  {author} {\bibinfo {author} {\bibfnamefont {K.~R.}\ \bibnamefont
  {Sreenivasan}},\ }\href {\doibase 10.1103/RevModPhys.71.S383} {\bibfield
  {journal} {\bibinfo  {journal} {Rev. Mod. Phys.}\ }\textbf {\bibinfo {volume}
  {71}},\ \bibinfo {pages} {S383} (\bibinfo {year} {1999})}\BibitemShut
  {NoStop}%
\bibitem [{\citenamefont {Ruddick}\ \emph {et~al.}(2000)\citenamefont
  {Ruddick}, \citenamefont {Anis},\ and\ \citenamefont
  {Thompson}}]{ruddick-2000}%
  \BibitemOpen
  \bibfield  {author} {\bibinfo {author} {\bibfnamefont {B.}~\bibnamefont
  {Ruddick}}, \bibinfo {author} {\bibfnamefont {A.}~\bibnamefont {Anis}}, \
  and\ \bibinfo {author} {\bibfnamefont {K.}~\bibnamefont {Thompson}},\ }\href
  {\doibase 10.1175/1520-0426(2000)017<1541:MLSFTB>2.0.CO;2} {\bibfield
  {journal} {\bibinfo  {journal} {Journal of Atmospheric and Oceanic
  Technology}\ }\textbf {\bibinfo {volume} {17}},\ \bibinfo {pages} {1541}
  (\bibinfo {year} {2000})}\BibitemShut {NoStop}%
\bibitem [{\citenamefont {{Vaughan, S.}}(2005)}]{vaughan-2005}%
  \BibitemOpen
  \bibfield  {author} {\bibinfo {author} {\bibnamefont {{Vaughan, S.}}},\
  }\href {\doibase 10.1051/0004-6361:20041453} {\bibfield  {journal} {\bibinfo
  {journal} {A\&A}\ }\textbf {\bibinfo {volume} {431}},\ \bibinfo {pages} {391}
  (\bibinfo {year} {2005})}\BibitemShut {NoStop}%
\bibitem [{\citenamefont {Bluteau}\ \emph {et~al.}(2011)\citenamefont
  {Bluteau}, \citenamefont {Jones},\ and\ \citenamefont {Ivey}}]{bluteau-2011}%
  \BibitemOpen
  \bibfield  {author} {\bibinfo {author} {\bibfnamefont {C.~E.}\ \bibnamefont
  {Bluteau}}, \bibinfo {author} {\bibfnamefont {N.~L.}\ \bibnamefont {Jones}},
  \ and\ \bibinfo {author} {\bibfnamefont {G.~N.}\ \bibnamefont {Ivey}},\
  }\href {\doibase 10.4319/lom.2011.9.302} {\bibfield  {journal} {\bibinfo
  {journal} {Limnology and Oceanography: Methods}\ }\textbf {\bibinfo {volume}
  {9}},\ \bibinfo {pages} {302} (\bibinfo {year} {2011})}\BibitemShut {NoStop}%
\bibitem [{\citenamefont {Choudhuri}\ \emph {et~al.}(2004)\citenamefont
  {Choudhuri}, \citenamefont {Ghosal},\ and\ \citenamefont
  {Roy}}]{choudhuri-2004}%
  \BibitemOpen
  \bibfield  {author} {\bibinfo {author} {\bibfnamefont {N.}~\bibnamefont
  {Choudhuri}}, \bibinfo {author} {\bibfnamefont {S.}~\bibnamefont {Ghosal}}, \
  and\ \bibinfo {author} {\bibfnamefont {A.}~\bibnamefont {Roy}},\ }\href
  {\doibase 10.1198/016214504000000557} {\bibfield  {journal} {\bibinfo
  {journal} {Journal of the American Statistical Association}\ }\textbf
  {\bibinfo {volume} {99}},\ \bibinfo {pages} {1050} (\bibinfo {year}
  {2004})}\BibitemShut {NoStop}%
\bibitem [{\citenamefont {Kraichnan}(1959)}]{kraichnan-1959}%
  \BibitemOpen
  \bibfield  {author} {\bibinfo {author} {\bibfnamefont {R.~H.}\ \bibnamefont
  {Kraichnan}},\ }\href {\doibase 10.1017/S0022112059000362} {\bibfield
  {journal} {\bibinfo  {journal} {Journal of Fluid Mechanics}\ }\textbf
  {\bibinfo {volume} {5}},\ \bibinfo {pages} {497–543} (\bibinfo {year}
  {1959})}\BibitemShut {NoStop}%
\bibitem [{\citenamefont {Sreenivasan}(1984)}]{sreenivasan-1984}%
  \BibitemOpen
  \bibfield  {author} {\bibinfo {author} {\bibfnamefont {K.~R.}\ \bibnamefont
  {Sreenivasan}},\ }\href {\doibase 10.1063/1.864731} {\bibfield  {journal}
  {\bibinfo  {journal} {The Physics of Fluids}\ }\textbf {\bibinfo {volume}
  {27}},\ \bibinfo {pages} {1048} (\bibinfo {year} {1984})}\BibitemShut
  {NoStop}%
\bibitem [{\citenamefont {Saddoughi}\ and\ \citenamefont
  {Veeravalli}(1994)}]{saddoughi-veeravalli-1994}%
  \BibitemOpen
  \bibfield  {author} {\bibinfo {author} {\bibfnamefont {S.~G.}\ \bibnamefont
  {Saddoughi}}\ and\ \bibinfo {author} {\bibfnamefont {S.~V.}\ \bibnamefont
  {Veeravalli}},\ }\href {\doibase 10.1017/S0022112094001370} {\bibfield
  {journal} {\bibinfo  {journal} {Journal of Fluid Mechanics}\ }\textbf
  {\bibinfo {volume} {268}},\ \bibinfo {pages} {333–372} (\bibinfo {year}
  {1994})}\BibitemShut {NoStop}%
\bibitem [{\citenamefont {Khurshid}\ \emph {et~al.}(2018)\citenamefont
  {Khurshid}, \citenamefont {Donzis},\ and\ \citenamefont
  {Sreenivasan}}]{khurshid-2018}%
  \BibitemOpen
  \bibfield  {author} {\bibinfo {author} {\bibfnamefont {S.}~\bibnamefont
  {Khurshid}}, \bibinfo {author} {\bibfnamefont {D.~A.}\ \bibnamefont
  {Donzis}}, \ and\ \bibinfo {author} {\bibfnamefont {K.~R.}\ \bibnamefont
  {Sreenivasan}},\ }\href {\doibase 10.1103/PhysRevFluids.3.082601} {\bibfield
  {journal} {\bibinfo  {journal} {Phys. Rev. Fluids}\ }\textbf {\bibinfo
  {volume} {3}},\ \bibinfo {pages} {082601} (\bibinfo {year}
  {2018})}\BibitemShut {NoStop}%
\bibitem [{\citenamefont {Bershadskii}(2019)}]{bershadskii-2019}%
  \BibitemOpen
  \bibfield  {author} {\bibinfo {author} {\bibfnamefont {A.}~\bibnamefont
  {Bershadskii}},\ }\href@noop {} {\enquote {\bibinfo {title} {Distributed
  chaos and turbulence in bénard-marangoni and rayleigh-bénard convection},}\
  } (\bibinfo {year} {2019}),\ \Eprint {http://arxiv.org/abs/1903.05018}
  {arXiv:1903.05018 [physics.flu-dyn]} \BibitemShut {NoStop}%
\bibitem [{\citenamefont {Contreras-Cristán}\ \emph
  {et~al.}(2006)\citenamefont {Contreras-Cristán}, \citenamefont
  {Gutiérrez-Peña},\ and\ \citenamefont {Walker}}]{contreras-cristan-2006}%
  \BibitemOpen
  \bibfield  {author} {\bibinfo {author} {\bibfnamefont {A.}~\bibnamefont
  {Contreras-Cristán}}, \bibinfo {author} {\bibfnamefont {E.}~\bibnamefont
  {Gutiérrez-Peña}}, \ and\ \bibinfo {author} {\bibfnamefont {S.~G.}\
  \bibnamefont {Walker}},\ }\href {\doibase 10.1080/03610910600880203}
  {\bibfield  {journal} {\bibinfo  {journal} {Communications in Statistics -
  Simulation and Computation}\ }\textbf {\bibinfo {volume} {35}},\ \bibinfo
  {pages} {857} (\bibinfo {year} {2006})}\BibitemShut {NoStop}%
\bibitem [{\citenamefont {Brillinger}(2001)}]{brillinger-2001}%
  \BibitemOpen
  \bibfield  {author} {\bibinfo {author} {\bibfnamefont {D.}~\bibnamefont
  {Brillinger}},\ }\href {https://books.google.com.au/books?id=PX5HExMKER0C}
  {\emph {\bibinfo {title} {Time Series: Data Analysis and Theory}}},\ Classics
  in Applied Mathematics\ (\bibinfo  {publisher} {Society for Industrial and
  Applied Mathematics},\ \bibinfo {year} {2001})\BibitemShut {NoStop}%
\bibitem [{\citenamefont {Whittle}(1957)}]{whittle-1957}%
  \BibitemOpen
  \bibfield  {author} {\bibinfo {author} {\bibfnamefont {P.}~\bibnamefont
  {Whittle}},\ }\href@noop {} {\bibfield  {journal} {\bibinfo  {journal}
  {Journal of the Royal Statistical Society}\ }\textbf {\bibinfo {volume}
  {19}},\ \bibinfo {pages} {38} (\bibinfo {year} {1957})}\BibitemShut {NoStop}%
\bibitem [{\citenamefont {Massey}(1951)}]{massey-1951}%
  \BibitemOpen
  \bibfield  {author} {\bibinfo {author} {\bibfnamefont {F.~J.}\ \bibnamefont
  {Massey}},\ }\href {\doibase 10.1080/01621459.1951.10500769} {\bibfield
  {journal} {\bibinfo  {journal} {Journal of the American Statistical
  Association}\ }\textbf {\bibinfo {volume} {46}},\ \bibinfo {pages} {68}
  (\bibinfo {year} {1951})}\BibitemShut {NoStop}%
\bibitem [{\citenamefont {Mikaelian}(1989)}]{mikaelian-1989}%
  \BibitemOpen
  \bibfield  {author} {\bibinfo {author} {\bibfnamefont {K.~O.}\ \bibnamefont
  {Mikaelian}},\ }\href {\doibase https://doi.org/10.1016/0167-2789(89)90089-4}
  {\bibfield  {journal} {\bibinfo  {journal} {Physica D: Nonlinear Phenomena}\
  }\textbf {\bibinfo {volume} {36}},\ \bibinfo {pages} {343 } (\bibinfo {year}
  {1989})}\BibitemShut {NoStop}%
\bibitem [{\citenamefont {Zhou}(2001)}]{zhou-2001}%
  \BibitemOpen
  \bibfield  {author} {\bibinfo {author} {\bibfnamefont {Y.}~\bibnamefont
  {Zhou}},\ }\href {\doibase 10.1063/1.1336151} {\bibfield  {journal} {\bibinfo
   {journal} {Physics of Fluids}\ }\textbf {\bibinfo {volume} {13}},\ \bibinfo
  {pages} {538} (\bibinfo {year} {2001})}\BibitemShut {NoStop}%
\bibitem [{\citenamefont {Obukhov}\ and\ \citenamefont
  {Yaglom}(1959)}]{obukhov-1959}%
  \BibitemOpen
  \bibfield  {author} {\bibinfo {author} {\bibfnamefont {A.~M.}\ \bibnamefont
  {Obukhov}}\ and\ \bibinfo {author} {\bibfnamefont {A.~M.}\ \bibnamefont
  {Yaglom}},\ }\href {\doibase 10.1002/qj.49708536402} {\bibfield  {journal}
  {\bibinfo  {journal} {Quarterly Journal of the Royal Meteorological Society}\
  }\textbf {\bibinfo {volume} {85}},\ \bibinfo {pages} {81} (\bibinfo {year}
  {1959})}\BibitemShut {NoStop}%
\bibitem [{\citenamefont {Bolgiano~Jr.}(1959)}]{bolgiano-1959}%
  \BibitemOpen
  \bibfield  {author} {\bibinfo {author} {\bibfnamefont {R.}~\bibnamefont
  {Bolgiano~Jr.}},\ }\href {\doibase 10.1029/JZ064i012p02226} {\bibfield
  {journal} {\bibinfo  {journal} {Journal of Geophysical Research (1896-1977)}\
  }\textbf {\bibinfo {volume} {64}},\ \bibinfo {pages} {2226} (\bibinfo {year}
  {1959})}\BibitemShut {NoStop}%
\bibitem [{\citenamefont {Batchelor}(1969)}]{batchelor-1969}%
  \BibitemOpen
  \bibfield  {author} {\bibinfo {author} {\bibfnamefont {G.~K.}\ \bibnamefont
  {Batchelor}},\ }\href {\doibase 10.1063/1.1692443} {\bibfield  {journal}
  {\bibinfo  {journal} {The Physics of Fluids}\ }\textbf {\bibinfo {volume}
  {12}},\ \bibinfo {pages} {II} (\bibinfo {year} {1969})}\BibitemShut {NoStop}%
\bibitem [{\citenamefont {Akula}(2014)}]{akula-thesis}%
  \BibitemOpen
  \bibfield  {author} {\bibinfo {author} {\bibfnamefont {B.}~\bibnamefont
  {Akula}},\ }\emph {\bibinfo {title} {Experimental Investigation of Buoyancy
  Driven Mixing With and Without Shear at Different Atwood Numbers}},\
  \href@noop {} {Ph.D. thesis},\ \bibinfo  {school} {Texas A \& M University}
  (\bibinfo {year} {2014})\BibitemShut {NoStop}%
\bibitem [{\citenamefont {Kraft}\ \emph {et~al.}(2009)\citenamefont {Kraft},
  \citenamefont {Banerjee},\ and\ \citenamefont {Andrews}}]{kraft-2009}%
  \BibitemOpen
  \bibfield  {author} {\bibinfo {author} {\bibfnamefont {W.~N.}\ \bibnamefont
  {Kraft}}, \bibinfo {author} {\bibfnamefont {A.}~\bibnamefont {Banerjee}}, \
  and\ \bibinfo {author} {\bibfnamefont {M.~J.}\ \bibnamefont {Andrews}},\
  }\href {\doibase 10.1007/s00348-009-0636-3} {\bibfield  {journal} {\bibinfo
  {journal} {Experiments in Fluids}\ }\textbf {\bibinfo {volume} {47}},\
  \bibinfo {pages} {49} (\bibinfo {year} {2009})}\BibitemShut {NoStop}%
\bibitem [{\citenamefont {Taylor}(1950)}]{taylor-1950}%
  \BibitemOpen
  \bibfield  {author} {\bibinfo {author} {\bibfnamefont {G.~I.}\ \bibnamefont
  {Taylor}},\ }\href {\doibase 10.1098/rspa.1950.0050} {\bibfield  {journal}
  {\bibinfo  {journal} {Proceedings of the Royal Society of London}\ }\textbf
  {\bibinfo {volume} {201}},\ \bibinfo {pages} {175} (\bibinfo {year}
  {1950})}\BibitemShut {NoStop}%
\bibitem [{\citenamefont {Ryutov}\ \emph {et~al.}(2000)\citenamefont {Ryutov},
  \citenamefont {Derzon},\ and\ \citenamefont {Matzen}}]{Ryutov}%
  \BibitemOpen
  \bibfield  {author} {\bibinfo {author} {\bibfnamefont {D.~D.}\ \bibnamefont
  {Ryutov}}, \bibinfo {author} {\bibfnamefont {M.~S.}\ \bibnamefont {Derzon}},
  \ and\ \bibinfo {author} {\bibfnamefont {M.~K.}\ \bibnamefont {Matzen}},\
  }\href {\doibase 10.1103/RevModPhys.72.167} {\bibfield  {journal} {\bibinfo
  {journal} {Rev. Mod. Phys.}\ }\textbf {\bibinfo {volume} {72}},\ \bibinfo
  {pages} {167} (\bibinfo {year} {2000})}\BibitemShut {NoStop}%
\bibitem [{\citenamefont {Gekelman}\ \emph {et~al.}(2014)\citenamefont
  {Gekelman}, \citenamefont {Compernolle}, \citenamefont {DeHaas},\ and\
  \citenamefont {Vincena}}]{gekelman-2014}%
  \BibitemOpen
  \bibfield  {author} {\bibinfo {author} {\bibfnamefont {W.}~\bibnamefont
  {Gekelman}}, \bibinfo {author} {\bibfnamefont {B.~V.}\ \bibnamefont
  {Compernolle}}, \bibinfo {author} {\bibfnamefont {T.}~\bibnamefont {DeHaas}},
  \ and\ \bibinfo {author} {\bibfnamefont {S.}~\bibnamefont {Vincena}},\ }\href
  {http://stacks.iop.org/0741-3335/56/i=6/a=064002} {\bibfield  {journal}
  {\bibinfo  {journal} {Plasma Physics and Controlled Fusion}\ }\textbf
  {\bibinfo {volume} {56}},\ \bibinfo {pages} {064002} (\bibinfo {year}
  {2014})}\BibitemShut {NoStop}%
\bibitem [{\citenamefont
  {{Kolmogorov}}(1941{\natexlab{c}})}]{kolmogorov-1941c}%
  \BibitemOpen
  \bibfield  {author} {\bibinfo {author} {\bibfnamefont {A.}~\bibnamefont
  {{Kolmogorov}}},\ }\href@noop {} {\bibfield  {journal} {\bibinfo  {journal}
  {Dokl. Akad. Nauk SSSR}\ }\textbf {\bibinfo {volume} {32}},\ \bibinfo {pages}
  {19} (\bibinfo {year} {1941}{\natexlab{c}})}\BibitemShut {NoStop}%
\bibitem [{\citenamefont {Kolmogorov}(1991)}]{kolmogorov-1991}%
  \BibitemOpen
  \bibfield  {author} {\bibinfo {author} {\bibfnamefont {A.~N.}\ \bibnamefont
  {Kolmogorov}},\ }\href@noop {} {\bibfield  {journal} {\bibinfo  {journal}
  {Proceedings: Mathematical and Physical Sciences}\ }\textbf {\bibinfo
  {volume} {434}},\ \bibinfo {pages} {9} (\bibinfo {year} {1991})}\BibitemShut
  {NoStop}%
\bibitem [{\citenamefont {Bandt}\ and\ \citenamefont
  {Pompe}(2002)}]{bandt-pompe-2002}%
  \BibitemOpen
  \bibfield  {author} {\bibinfo {author} {\bibfnamefont {C.}~\bibnamefont
  {Bandt}}\ and\ \bibinfo {author} {\bibfnamefont {B.}~\bibnamefont {Pompe}},\
  }\href {\doibase 10.1103/PhysRevLett.88.174102} {\bibfield  {journal}
  {\bibinfo  {journal} {Phys. Rev. Lett.}\ }\textbf {\bibinfo {volume} {88}},\
  \bibinfo {pages} {174102} (\bibinfo {year} {2002})}\BibitemShut {NoStop}%
\bibitem [{\citenamefont {Rosso}\ \emph {et~al.}(2007)\citenamefont {Rosso},
  \citenamefont {Larrondo}, \citenamefont {Martin}, \citenamefont {Plastino},\
  and\ \citenamefont {Fuentes}}]{rosso-2007}%
  \BibitemOpen
  \bibfield  {author} {\bibinfo {author} {\bibfnamefont {O.~A.}\ \bibnamefont
  {Rosso}}, \bibinfo {author} {\bibfnamefont {H.~A.}\ \bibnamefont {Larrondo}},
  \bibinfo {author} {\bibfnamefont {M.~T.}\ \bibnamefont {Martin}}, \bibinfo
  {author} {\bibfnamefont {A.}~\bibnamefont {Plastino}}, \ and\ \bibinfo
  {author} {\bibfnamefont {M.~A.}\ \bibnamefont {Fuentes}},\ }\href {\doibase
  10.1103/PhysRevLett.99.154102} {\bibfield  {journal} {\bibinfo  {journal}
  {Phys. Rev. Lett.}\ }\textbf {\bibinfo {volume} {99}},\ \bibinfo {pages}
  {154102} (\bibinfo {year} {2007})}\BibitemShut {NoStop}%
\bibitem [{\citenamefont {Maggs}\ and\ \citenamefont
  {Morales}(2013)}]{maggs-morales-2013}%
  \BibitemOpen
  \bibfield  {author} {\bibinfo {author} {\bibfnamefont {J.~E.}\ \bibnamefont
  {Maggs}}\ and\ \bibinfo {author} {\bibfnamefont {G.~J.}\ \bibnamefont
  {Morales}},\ }\href {http://stacks.iop.org/0741-3335/55/i=8/a=085015}
  {\bibfield  {journal} {\bibinfo  {journal} {Plasma Physics and Controlled
  Fusion}\ }\textbf {\bibinfo {volume} {55}},\ \bibinfo {pages} {085015}
  (\bibinfo {year} {2013})}\BibitemShut {NoStop}%
\bibitem [{\citenamefont {López-Ruiz}\ \emph {et~al.}(1995)\citenamefont
  {López-Ruiz}, \citenamefont {Mancini},\ and\ \citenamefont
  {Calbet}}]{lopezruiz-1995}%
  \BibitemOpen
  \bibfield  {author} {\bibinfo {author} {\bibfnamefont {R.}~\bibnamefont
  {López-Ruiz}}, \bibinfo {author} {\bibfnamefont {H.}~\bibnamefont
  {Mancini}}, \ and\ \bibinfo {author} {\bibfnamefont {X.}~\bibnamefont
  {Calbet}},\ }\href {\doibase https://doi.org/10.1016/0375-9601(95)00867-5}
  {\bibfield  {journal} {\bibinfo  {journal} {Physics Letters A}\ }\textbf
  {\bibinfo {volume} {209}},\ \bibinfo {pages} {321 } (\bibinfo {year}
  {1995})}\BibitemShut {NoStop}%
\bibitem [{\citenamefont {Martin}\ \emph {et~al.}(2006)\citenamefont {Martin},
  \citenamefont {Plastino},\ and\ \citenamefont {Rosso}}]{martin-2006}%
  \BibitemOpen
  \bibfield  {author} {\bibinfo {author} {\bibfnamefont {M.}~\bibnamefont
  {Martin}}, \bibinfo {author} {\bibfnamefont {A.}~\bibnamefont {Plastino}}, \
  and\ \bibinfo {author} {\bibfnamefont {O.}~\bibnamefont {Rosso}},\ }\href
  {\doibase https://doi.org/10.1016/j.physa.2005.11.053} {\bibfield  {journal}
  {\bibinfo  {journal} {Physica A: Statistical Mechanics and its Applications}\
  }\textbf {\bibinfo {volume} {369}},\ \bibinfo {pages} {439 } (\bibinfo {year}
  {2006})}\BibitemShut {NoStop}%
\bibitem [{\citenamefont {Bauke}(2007)}]{bauke-2007}%
  \BibitemOpen
  \bibfield  {author} {\bibinfo {author} {\bibfnamefont {H.}~\bibnamefont
  {Bauke}},\ }\href {\doibase 10.1140/epjb/e2007-00219-y} {\bibfield  {journal}
  {\bibinfo  {journal} {The European Physical Journal B}\ }\textbf {\bibinfo
  {volume} {58}},\ \bibinfo {pages} {167} (\bibinfo {year} {2007})}\BibitemShut
  {NoStop}%
\bibitem [{\citenamefont {Buehler}\ \emph {et~al.}(2007)\citenamefont
  {Buehler}, \citenamefont {Tang}, \citenamefont {van Duin},\ and\
  \citenamefont {Goddard}}]{buehler-2007}%
  \BibitemOpen
  \bibfield  {author} {\bibinfo {author} {\bibfnamefont {M.~J.}\ \bibnamefont
  {Buehler}}, \bibinfo {author} {\bibfnamefont {H.}~\bibnamefont {Tang}},
  \bibinfo {author} {\bibfnamefont {A.~C.~T.}\ \bibnamefont {van Duin}}, \ and\
  \bibinfo {author} {\bibfnamefont {W.~A.}\ \bibnamefont {Goddard}},\ }\href
  {\doibase 10.1103/PhysRevLett.99.165502} {\bibfield  {journal} {\bibinfo
  {journal} {Phys. Rev. Lett.}\ }\textbf {\bibinfo {volume} {99}},\ \bibinfo
  {pages} {165502} (\bibinfo {year} {2007})}\BibitemShut {NoStop}%
\bibitem [{\citenamefont {Rana}\ and\ \citenamefont
  {Herrmann}(2011)}]{rana-2011}%
  \BibitemOpen
  \bibfield  {author} {\bibinfo {author} {\bibfnamefont {S.}~\bibnamefont
  {Rana}}\ and\ \bibinfo {author} {\bibfnamefont {M.}~\bibnamefont
  {Herrmann}},\ }\href {\doibase 10.1063/1.3640022} {\bibfield  {journal}
  {\bibinfo  {journal} {Physics of Fluids}\ }\textbf {\bibinfo {volume} {23}},\
  \bibinfo {pages} {091109} (\bibinfo {year} {2011})}\BibitemShut {NoStop}%
\bibitem [{\citenamefont {{Neuvazhaev}}(1975)}]{neuvazhaev-1975}%
  \BibitemOpen
  \bibfield  {author} {\bibinfo {author} {\bibfnamefont {V.~E.}\ \bibnamefont
  {{Neuvazhaev}}},\ }\href@noop {} {\bibfield  {journal} {\bibinfo  {journal}
  {Soviet Physics Doklady}\ }\textbf {\bibinfo {volume} {20}},\ \bibinfo
  {pages} {1053} (\bibinfo {year} {1975})}\BibitemShut {NoStop}%
\end{thebibliography}%

%\appendix

\end{document}